\documentclass[fleqn,usenatbib]{mnras}

\usepackage[T1]{fontenc}

\DeclareRobustCommand{\VAN}[3]{#2}
\let\VANthebibliography\thebibliography
\def\thebibliography{\DeclareRobustCommand{\VAN}[3]{##3}\VANthebibliography}

\usepackage{graphicx}	
\usepackage{amsmath}	
\usepackage{amssymb}	
\usepackage{bm}
\usepackage{multirow}
\usepackage{booktabs}

\usepackage{pdflscape}
\usepackage{color}
\usepackage{threeparttable}

\title[Subhalo collision and formation of dwarf galaxy]{
Frequency of the dark matter subhalo collisions and bifurcation sequence arising formation of dwarf galaxies
}

\author[K. Otaki and M. Mori]{
Koki Otaki$^{1}$\thanks{E-mail: otaki@ccs.tsukuba.ac.jp} and 
Masao Mori$^{2}$
\\
$^{1}$Graduate School of Pure and Applied Sciences, University of Tsukuba, Tennodai 1-1-1, Tsukuba, Ibaraki 305-8577, Japan\\
$^{2}$Center for Computational Sciences, University of Tsukuba, Tennodai 1-1-1, Tsukuba, Ibaraki 305-8577, Japan
}

\date{Accepted XXX. Received YYY; in original form ZZZ}

\pubyear{2021}

\begin{document}
\label{firstpage}
\pagerange{\pageref{firstpage}--\pageref{lastpage}}
\maketitle

\begin{abstract}
The cold dark matter (CDM) model predicts galaxies have 100 times more dark matter mass than stars. Nevertheless, recent observations report the existence of dark-matter-deficient galaxies with less dark matter than expected. 
To solve this problem, 
we investigate the physical processes of galaxy formation in head-on collisions between gas-containing dark matter subhaloes (DMSHs). 
Analytical estimation of the collision frequency between DMSHs associated with a massive host halo indicates that collisions frequently occur within 1/10th of the virial radius of the host halo, with a collision timescale of about $10\, \mathrm{Myr}$, and the most frequent relative velocity increases with increasing radius.
Using analytical models and numerical simulations, we show the bifurcation channel of the formation of dark-matter-dominated and dark-matter-deficient galaxies.
In the case of low-velocity collisions, a dark-matter-dominated galaxy is formed by the merging of two DMSHs.  
In the case of moderate-velocity collisions, the two DMSHs penetrate each other. However the gas medium collides, and star formation begins as the gas density increases, forming a dwarf galaxy without dark matter at the collision surface.
In the case of high-velocity collisions, shock-breakout occurs due to the shock waves generated at the collision surface reaching the gas surface, and no galaxy forms.
For example, the simulation demonstrates that a pair of DMSHs with a mass of $10^9\,\mathrm{M_\odot}$ containing gas of 0.1 solar metallicity 
forms a dark-matter-deficient galaxy with a stellar mass of $10^7\,\mathrm{M_\odot}$ 
for a relative velocity of $200\, \mathrm{km\,s^{-1}}$. 
\end{abstract}




\section{Introduction}

Cold dark matter (CDM) drives the hierarchical structure formation in the standard galaxy formation model. In other words, cosmic structures are believed to form in a bottom-up fashion in which small dark matter haloes repeatedly collide and merge, growing into larger systems.

While the CDM model successfully reproduces statistical properties such as the large--scale distribution of galaxies in the universe, some serious inconsistencies exist between theoretical predictions and observations on a few Mpc or less
\citep{MooreGhignaGovernatoLakeEtAl_1999_DarkMatterSubstructure_ApJ, KlypinKravtsovValenzuelaPrada_1999_WhereAreMissing_ApJ}. 

The number of satellite galaxies observed around the Milky Way is more than one order of magnitude less than that of dark matter subhaloes (DMSHs) predicted by the CDM model.
This discrepancy is known as the missing satellite problem. 
It implies that there may be a huge number of extremely faint galaxies and dark matter-dominated haloes with little or no stellar component in the Local Group.

From a general perspective, the correlation between stellar components and dark matter haloes has been of profound interest and hitherto numerous studies from both theoretical and observational viewpoints.
Almost all studies about the relationship between the stellar mass and the dark matter halo mass in galaxies have shown that the dark matter fraction in galaxies is expected to be more than 90\%
\cite[e.g., ][]{BehrooziWechslerConroy_2013_AVERAGESTARFORMATION_ApJ}.

However, \cite{vanDokkumDanieliCohenMerrittEtAl_2018_GalaxyLackingDark_Nature} 
recently reported that the satellite galaxy NGC1052-DF2, a member of the elliptical galaxy NGC1052 group, has very little dark matter component compared to the theoretical predictions. 
Its stellar mass is $2\times 10^8\,\mathrm{M_\odot}$, whereas its dynamical mass is $<3.4\times 10^8\,\mathrm{M_\odot}$ within a radius of $7.6\,\mathrm{kpc}$. 
This radius is the position of the outermost globular clusters in this galaxy, which is greater than its effective radius of $2.2\,\mathrm{kpc}$. 
While the derivation of dynamical masses using velocity dispersion of globular clusters indicates serious ambiguity \cite[e.g., ][]{HayashiInoue_2018_EffectsMassModels_MNRAS}, the detailed analysis of the Jeans model shows the lack of dark matter in those galaxies \citep{WassermanRomanowskyBrodieDokkumEtAl_2018_DeficitDarkMatter_ApJLa}.

In addition, NGC1052-DF4 in the NGC1052 group has also been discovered as a galaxy with similar properties  \citep{DokkumDanieliAbrahamConroyEtAl_2019_SecondGalaxyMissing_ApJL}. Its stellar mass is $1.5\times 10^8\,\mathrm{M_\odot}$, and its dynamical mass is estimated as $0.4\times10^8\,\mathrm{M_\odot}$ within $7\,\mathrm{kpc}$ from the galaxy centre. These two galaxies are classified as ultra-diffuse galaxies (UDGs). UDGs are peculiar galaxies with extremely low surface brightness, $\mu(g,0)>24\,\mathrm{mag\,arcsec^{-2}}$, and large effective radii, $r_\mathrm{e}>1.5\,\mathrm{kpc}$, first discovered in the Coma Cluster \citep{DokkumAbrahamMerrittZhangEtAl_2015_FORTYSEVENMILKYWAYSIZED_ApJL, KodaYagiYamanoiKomiyama_2015_APPROXIMATHOUSANDULTRADIFFUSE_ApJL}. 
It should be noted that the issue of uncertainty in the distance to these galaxies continues to active discussion and does not yet converge
 \citep{TrujilloBeasleyBorlaffCarrascoEtAl_2019_Distance13Mpc_MNRAS,MonelliTrujillo_2019_TRGBDistanceSecond_ApJL}.
In the latest reports, \cite{DanieliDokkumAbrahamConroyEtAl_2020_TipRedGiant_ApJL} and \cite{ShenDanieliDokkumAbrahamEtAl_2021_TipRedGiant_ApJL} claim that the distances of NGC1052-DF2 and NGC1052-DF4 are $\sim20\,\mathrm{Mpc}$, which is derived by Hubble Space Telescope Advanced Camera for Surveys imaging, and that these two galaxies are dark-matter--deficient galaxies.



Furthermore, different observations have also reported the existence of other dark-matter-deficient galaxies.
\cite{ManceraPinaFraternaliAdamsMarascoEtAl_2019_BaryonicTullyFisher_ApJL,ManceraPinaFraternaliOmanAdamsEtAl_2020_RobustKinematicsGasrich_MNRAS} found six $\mathrm{H_{\,I}}$-rich UDGs which have a high baryon fraction within a radius larger than the effective radius. 
\cite{ManceraPinaFraternaliOosterlooAdamsEtAl_2022_NoNeedDark_MonthlyNoticesoftheRoyalAstronomicalSociety} observed the AGC 114905, one of the six $\mathrm{H_{\,I}}$-rich UDGs, at high spatial resolution using the Karl G. Jansky Very Large Array. The $\mathrm{H_{\,I}}$ rotation curve of the galaxy is fitted by the baryon contribution alone up to the observed outermost radius. This galaxy has the stellar mass of  $M_\star=(1.3\pm0.3)\times10^8\,\mathrm{M_\odot}$, and the $\mathrm{H_{\,I}}$ mass of  $M_\mathrm{H_{\,I}}=(9.7\pm1.4)\times10^8\,\mathrm{M_\odot}$.
\cite{GuoHuZhengLiaoEtAl_2020_FurtherEvidencePopulation_NatAstron} reported 19 dwarf galaxies that have high baryon fraction within $\mathrm{H_{\,I}}$ radius $r_\mathrm{H_{\,I}}$ , which is defined at $\mathrm{H_{\, I}}$ surface density $=1\,\mathrm{M_\odot\,pc^{-2}}$. This radius is larger than the effective radius. 
Some of these galaxies have large effective radii, which may be UDGs.
As one example, the galaxy AGC 213086, with $r_\mathrm{H_{\,I}}=14.37\pm1.023\,\mathrm{kpc}$, has a stellar mass of $5.51^{+4.02}_{-2.32}\times10^8\,\mathrm{M_\odot}$, an $\mathrm{H_{\, I}}$ mass of $2.45^{+0.11}_{-0.09}\times10^9\,\mathrm{M_\odot}$ and a dynamical mass of $6.31^{+0.89}_{-0.77}\times10^9\,\mathrm{M_\odot}$.
They mentioned that 14 of the 19 galaxies are isolated from any group of galaxies and are located in the field. 
Thus far, a total of 27 dark-matter-deficient galaxies have already been identified. However, it is still an open question to reveal the formation of dark-matter-deficient galaxies in the dark-matter-dominated universe.

There are several theoretical studies on the formation of dark-matter-deficient galaxies.
\cite{Ogiya_2018_TidalStrippingPossible_MNRAS} investigated the formation of dark-matter-deficient galaxies by tidal interaction between a host galaxy and a satellite galaxy using $N$-body simulations.
Assuming that the dark halo of the satellite galaxy has a cored density profile with a tightly bound and extremely radial orbit, the simulation results show that the effect of the tidal stripping successfully reproduces the observed properties of the NGC1052-DF2-like galaxies.
Similarly, \citet{YangYuAn_2020_SelfInteractingDarkMatter_Phys.Rev.Lett.b} demonstrate the formation of dark-matter-deficient galaxies driven by tidal interaction within the framework of a self-interacting dark matter (SIDM). So far, tidal interaction models have brought some success as a model to explain the formation of dark-matter-dominated galaxies \citep[see also][]{Nusser_2020_ScenarioUltradiffuseSatellite_ApJ}. However, \cite{MullerRichRomanYildizEtAl_2019_TidalTaleDetection_A&A} concluded from observations using the Jeanne Rich telescope that NGC1052-DF2 and NGC1052-DF4 have no evidence of tidal interaction. \cite{MontesTrujilloInfante-SainzMonelliEtAl_2021_DiskNoSignatures_ApJ} report that the stellar distribution of NGC1052-DF2 indicates no signatures of tidal distortion, but NGC1052-DF4 appears to have experienced tidal disruption \citep{MontesInfante-SainzMadrigal-AguadoRomanEtAl_2020_GalaxyMissingDark_ApJ}. Some dark-matter-deficient galaxies inhabit the fields and are not bound to a more massive host galaxy. In other words, they seem to be free from tidal stripping. 


Recently, it has been pointed out that high-velocity collisions between gas-rich dwarf galaxies are also capable of forming dark-matter-deficient galaxies \citep{Silk_2019_UltradiffuseGalaxiesDark_MNRAS, ShinJungKwonKimEtAl_2020_DarkMatterDeficient_ApJ, LeeShinKim_2021_DarkMatterDeficient_ApJL, OtakiMori_2022_FormationDarkmatterdeficientGalaxies_J.Phys.:Conf.Ser., OtakiMori_2023_CollisioninducedFormationDarkmatterdeficient_IAUSymp.}. 
\cite{Silk_2019_UltradiffuseGalaxiesDark_MNRAS}
advocates that they are scaled-down versions of Bullet cluster-like events and involve high-velocity collisions of gas-rich dwarf galaxies in high-density environments.
\cite{ShinJungKwonKimEtAl_2020_DarkMatterDeficient_ApJ} showed that dark-matter-deficient galaxies formed when two dwarf galaxies collide with each other at a relative velocity of $300\,\mathrm{km\,s^{-1}}$ using self-gravitating hydrodynamics simulations. 
They investigated the formation of dark-matter-deficient galaxies by running several simulations with various collision parameters, disk angles, mass ratios and gas fractions, as well as relative velocities of the two dwarf galaxies.
Furthermore, they utilized cosmological simulations IllustrisTNG \citep{NaimanPillepichSpringelRamirez-RuizEtAl_2018_FirstResultsIllustrisTNG_MonthlyNoticesoftheRoyalAstronomicalSociety, MarinacciVogelsbergerPakmorTorreyEtAl_2018_FirstResultsIllustrisTNG_MonthlyNoticesoftheRoyalAstronomicalSociety,PillepichNelsonHernquistSpringelEtAl_2018_FirstResultsIllustrisTNG_MonNotRAstronSoc, SpringelPakmorPillepichWeinbergerEtAl_2018_FirstResultsIllustrisTNG_MonthlyNoticesoftheRoyalAstronomicalSociety, NelsonPillepichSpringelWeinbergerEtAl_2018_FirstResultsIllustrisTNG_MonthlyNoticesoftheRoyalAstronomicalSociety, NelsonPillepichSpringelPakmorEtAl_2019_FirstResultsTNG50_MonthlyNoticesoftheRoyalAstronomicalSociety, PillepichNelsonSpringelPakmorEtAl_2019_FirstResultsTNG50_MonthlyNoticesoftheRoyalAstronomicalSociety} the occurrence of galaxy collisions that lead to the formation of dark-matter-deficient galaxies. 
The complexity of the physical phenomena involved in their cosmological simulations makes it difficult to understand the physical conditions for the formation of dark-matter-deficient galaxies. They conclude that no valid collision events were identified due to the numerical resolution.
From the observational point of view, \citet{vanDokkumShenKeimTrujillo-GomezEtAl_2022_TrailDarkmatterfreeGalaxies_Nature} have reported that the spatial distribution of NGC1052-DF2 and NGC1052-DF4 and their surrounding dwarf galaxies is due to the traces of collisions between dwarf galaxies \citep[see also,][]{vanDokkumShenRomanowskyAbrahamEtAl_2022_MonochromaticGlobularClusters_Astrophys.J., BuzzoForbesBrodieJanssensEtAl_2023_LargescaleStructureGlobular_Mon.Not.R.Astron.Soc.}


On the other hand, Otaki et al. (2023) recently analysed the data set of the latest high-resolution cosmological simulation Phi-4096 presented by \citet{IshiyamaPradaKlypinSinhaEtAl_2021_UchuuSimulationsData_MonthlyNoticesoftheRoyalAstronomicalSociety}. They found that sub-galactic dark matter haloes frequently collide with each other at various relative velocities and concluded that galaxy collisions should considerably contribute to the formation of dark-matter-deficient galaxies.
In this context, it is essential to evaluate the frequency of collisions between DMSHs associated with more massive galaxies, independent of the method and resolution of numerical simulations, and analytical estimates such as those described in this paper will yield important insights.
From observational viewpoints, several recent studies of nearby galaxies have reported the discovery of faint structures indicating interactions between dwarf galaxies \citep{StierwaltBeslaPattonJohnsonEtAl_2015_TiNyTITANSROLE_ApJ}. 
For example, \citet{PaudelSmithYoonCalderon-CastilloEtAl_2018_CatalogMergingDwarf_ApJS} classified 177 dwarf galaxies of $<10^{10}\,\mathrm{M_\odot}$ with features of dwarf-dwarf interactions: interacting pairs, shell and tidal tail. \citet{PoulainMarleauHabasDucEtAl_2022_HIObservationsMATLAS_A&A} reported 12 dwarf merger candidates detected $\mathrm{H_{\, I}}$ line. 
In \citet{ChhatkuliPaudelBachchanAryalEtAl_2023_FormingBlueCompact_MonthlyNoticesoftheRoyalAstronomicalSociety}, an analysis of recent observational data shows that the formation of blue compact galaxies is linked to galaxy collisions between dwarf galaxies with a burst of star formation.
These observational facts allow us to infer that dwarf galaxy collisions are not rare but occur relatively frequently in the nearby universe. Furthermore, taking into account the missing satellite problem mentioned before, it is easy to imagine that dark matter sub-halo collisions are even more frequent than dwarf galaxy collisions in the dark side of the universe.

So far, theoretical studies always assume high-speed galaxy collision. However, we consider low-speed collision, in which two sub-galactic haloes will merge into one halo and form a dark-matter-dominated galaxy, is also important. It would be of great interest to understand which physical processes play an essential role in the bifurcation between dark-matter-dominated and dark-matter-deficient galaxies at their formation epoch through galaxy collision simulations under idealised conditions.

These situations motivate us to explore the head-on collision between DMSHs, investigating the physical conditions for forming dark matter-deficient-galaxies and the relationship between formation probability and collision frequency.

In this study, we focus on the head-on collision process between DMSHs and investigate the possibility of the formation of dark-matter-deficient galaxies by our original simulation code assuming a flat $\Lambda$CDM cosmology with $\Omega_\mathrm{m}$=0.315,\,$\Omega_\mathrm{b}=0.048,\,h=0.674$ in the Planck Collaboration (\citeyear{AghanimAkramiAshdownAumontEtAl_2020_Planck2018Results_A&Aa}). This paper is organised as follows.
Section 2 assesses the frequency of mutual collisions in subhaloes associated with a more massive dark matter halo having a Navarro--Frenk--White (NFW) density profile \citet{NavarroFrenkWhite_1996_StructureColdDark_Astrophys.J., NavarroFrenkWhite_1997_UniversalDensityProfile_ApJa}. Section 3 analyses the physical conditions for bifurcation channels between forming dark-matter-dominated galaxies and dark-matter-deficient galaxies by head-on collisions of such DMSHs using a simple one-dimensional hydrodynamic model. 
In Section 4, we present our numerical method for DMSH collisions incorporating star formation and supernova feedback in a hybrid three-dimensional hydrodynamic and $N$-body model. Subsequently, Section 5 describes the results of the simulations. In Section 6, we summarise the conclusion of this paper and devote a discussion of the limitations of our model and a comparison with previous studies. 


\section{Collision frequency between dark matter subhaloes}


We estimate the number of collisions between DMSHs moving within the virial radius of the host halo under dynamical equilibrium.

Here, we assume the velocity of DMSH follows to the velocity distribution function of the host halo.

The energy of a DMSH moving with velocity $v$ in the gravitational potential $\Phi_\mathrm{NFW}$ of the NFW profile generated by a  host halo with mass $M_\mathrm{host}$ is
\begin{gather}
    E=\frac12 v^2+\Phi_\mathrm{NFW}(r).
\end{gather}
Since we consider a DMSH bound to the host galaxy, $E$ is negative at all times. An NFW potential is given by
\begin{align}
    \Phi_\mathrm{NFW}(r) = -\frac{GM_\mathrm{host}}{R_\mathrm{200,\,host}}&\frac{c_\mathrm{host}}{\ln{(1+c_\mathrm{host})}-c_\mathrm{host}/(1+c_\mathrm{host})}\nonumber\\
    &\times \frac{\ln{(1+x)}}{x},\label{eq: NFWpot}
\end{align}
where 
\begin{align}
    x = \frac{r}{r_\mathrm{s,\,host}},\quad R_\mathrm{200,\,host}=\left(\frac{3M_\mathrm{host}}{4\pi \rho_{200}}\right)^{1/3},
\end{align}
$G$ is the gravitational constant, 

$c_\mathrm{host}=R_\mathrm{200,\,host}/r_\mathrm{s,\,host}$ is the concentration, $r_\mathrm{s,\,host}$ is a scale radius of a host halo, and $\rho_{200}$ is $200$ times the critical density of the universe.
So far, the studies demonstrate that the concentrations $c$ tightly correlate with the mass of host haloes $M_\mathrm{host}$ such as the $c\text{--}M$ relation \citep[e.g., ][]{BullockKolattSigadSomervilleEtAl_2001_ProfilesDarkHaloes_MonthlyNoticesoftheRoyalAstronomicalSociety,PradaKlypinCuestaBetancort-RijoEtAl_2012_HaloConcentrationsStandard_MonNotRAstronSoc, IshiyamaAndo_2020_AbundanceStructureSubhaloes_MonthlyNoticesoftheRoyalAstronomicalSociety, IshiyamaPradaKlypinSinhaEtAl_2021_UchuuSimulationsData_MonthlyNoticesoftheRoyalAstronomicalSociety}.
In the following, the energy and potential are expressed using the positive values, $\mathcal{E}=-E\text{ and }\Psi=-\Phi_\mathrm{NFW}$, respectively.

The distribution function given a spherical density profile can be calculated by Eddington's formula \citep{BinneyTremaine_2008_GalacticDynamicsSecond_GalacticDynamics:SecondEditionbyJamesBinneyandScottTremaine.ISBN978-0-691-13026-2HB.PublishedbyPrincetonUniversityPressPrincetonNJUSA2008.},
\begin{gather}
    f(\mathcal{E})=\frac{1}{\sqrt{8}\pi^2}\left[\frac{1}{\sqrt{\mathcal{E}}}\left(\frac{\mathrm{d}\nu}{\mathrm{d}\Psi}\right)_{\Psi=0}+\int^{\mathcal{E}}_0\frac{\mathrm{d}^2\nu}{\mathrm{d}\Psi^2}\frac{\mathrm{d}\Psi}{\sqrt{\mathcal{E}-\Psi}}\right],
\end{gather}
where $\nu$ is the probability density distribution of an NFW profile,
\begin{gather}
    \nu(r)=\frac{\rho_\mathrm{NFW}(r)}{M_\mathrm{host}}=\frac{g(c_\mathrm{host})}{4\pi R_\mathrm{200,\,host}^3}\frac{1}{x(1+x)^2},\\
    g(c)=\frac{c^3}{\ln(1+c)-c/(1+c)}.
\end{gather}

We assumed the distribution of DMSHs follows the distribution function $f(\mathcal{E})$ calculated from a host halo to simplify the motion of DMSHs.
It should be noted that although we assumed above that the host halo and the system of DMSHs are in a state of perfect relaxation, it is not trivial.
The velocity distribution function of DMSHs at position $r$ from the centre of a host halo becomes
\begin{gather}
    P_r(\bm{v})=\frac{f(\mathcal{E})}{\nu(r)},
\end{gather}
and for a system with an isotropic velocity,
\begin{gather}
    P_r(v)=4\pi v^2\frac{f(\mathcal{E})}{\nu(r)}.
\end{gather}

\begin{figure}
    \centering
    \includegraphics[width=\columnwidth]{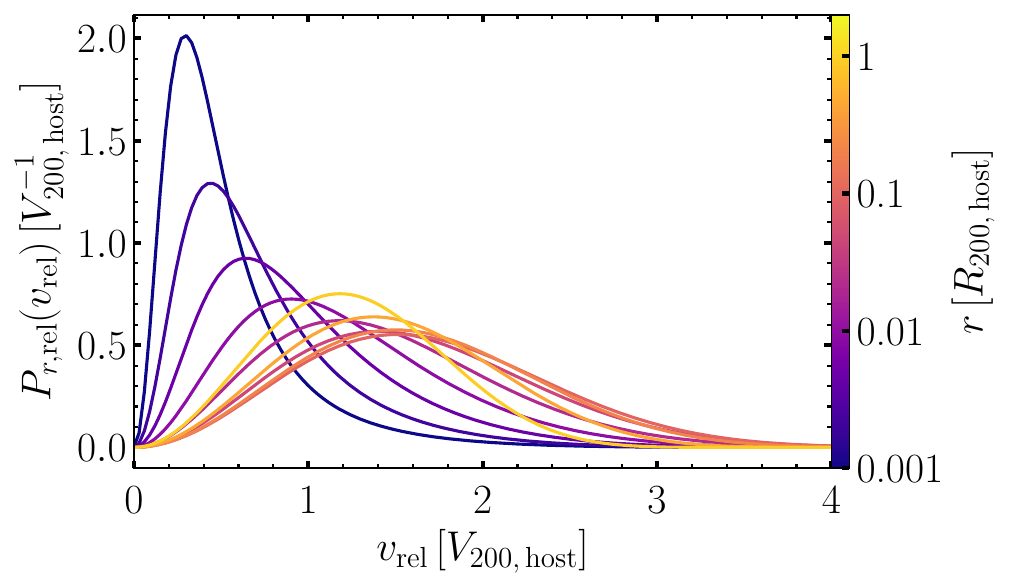}
    \caption{Probability distribution of the relative velocity for $c = 7.5$ calculated by equation \eqref{eq: Prrel}.}
    \label{fig: Prrel}
\end{figure}

Next, we solve the two-body problem for the distribution function \citep{FerrerHunter_2013_ImpactPhasespaceDensity_J.Cosmol.Astropart.Phys.}. 
The velocities of the two DMSHs are $\bm{v}_1,\text{ and }\bm{v}_2$.
Here, the DMSH is assumed to move following the velocity distribution function of the host halo in the NFW density distribution.
The velocity of the centre of mass and relative velocity are denoted as $\bm{v}_\mathrm{cm}=(\bm{v}_1+\bm{v}_2)/2$ and $\bm{v}_\mathrm{rel}=\bm{v}_1-\bm{v}_2$, respectively.
The velocity distribution function of the two DMSHs can be expressed in terms of the distribution function of the relative velocity and the velocity of the centre of mass: 
\begin{align}
    P_r(\bm{v}_1)&P_r(\bm{v}_2)\mathrm{d}^3\bm{v}_1\mathrm{d}^3\bm{v}_2\nonumber\\
    =&P_r(\bm{v}_\mathrm{cm}+\bm{v}_\mathrm{rel}/2)P_{r}(\bm{v}_\mathrm{cm}-\bm{v}_\mathrm{rel}/2)\mathrm{d}^3\bm{v}_\mathrm{cm}\mathrm{d} ^3\bm{v}_\mathrm{rel}. \label{eq: two-body}
\end{align}
The probability distribution of relative velocity $P_{r,\mathrm{rel}}$ at position $r$ is integrated only over the velocity of the centre of mass: 
\begin{gather}
    P_{r,\mathrm{rel}}(\bm{v}_\mathrm{rel}) = \int P_{r}(\bm{v}_\mathrm{cm}+\bm{v}_\mathrm{rel}/2)P_{r}(\bm{v}_\mathrm{cm}-\bm{v}_\mathrm{rel}/2)\mathrm{d}^3\bm{v}_\mathrm{cm},
\end{gather}
and for the case of isotropic velocity,
\begin{gather}
    P_{r,\mathrm{rel}}(v_\mathrm{rel}) 
    = \frac{8\pi^2v_\mathrm{rel}^2}{\nu(r)^2}\int_0^\infty\mathrm{d}{v}_\mathrm{cm} v_\mathrm{cm}^2 \int_{-1}^{1}\mathrm{d} z \,f(\mathcal{E}_1)f(\mathcal{E}_2),\label{eq: Prrel} \\
    \mathcal{E}_1 = -\frac12(v_\mathrm{cm}^2+v_\mathrm{rel}^2/4+v_\mathrm{cm}v_\mathrm{rel}z) + \Psi(r)\\
    \mathcal{E}_2 = -\frac12(v_\mathrm{cm}^2+v_\mathrm{rel}^2/4-v_\mathrm{cm}v_\mathrm{rel}z) + \Psi(r)
\end{gather}
where $v_\mathrm{cm}=|\bm{v}_\mathrm{cm}|,\,v_\mathrm{rel}=|\bm{v}_\mathrm{rel}|,\text{ and }\bm{v}_\mathrm{cm}\cdot\bm{v}_\mathrm{rel}=v_\mathrm{cm}v_\mathrm{rel}z$.
Fig. \ref{fig: Prrel} shows the probability distribution of relative velocity $P_{r,\mathrm{rel}}$ corresponding to each position $r$ from $0.001\,R_\mathrm{200\,host}$ to $2\,R_\mathrm{200,\,host}$ for $c=7.5$, which corresponds to the host mass $M_\mathrm{host}=10^{12}\,\mathrm{M_\odot}$ for the $c\text{--}M$ relation \citep{PradaKlypinCuestaBetancort-RijoEtAl_2012_HaloConcentrationsStandard_MonNotRAstronSoc}. The horizontal axis corresponds to the relative velocity between two DMSHs, normalised by the circular velocity $V_\mathrm{200,\,host}=\sqrt{GM_\mathrm{200,\,host}/R_\mathrm{200,\,host}}$.

Here, we define the expected value of relative velocities as
\begin{gather}
    E[v_{r,\mathrm{rel}}] = \int \mathrm{d}v_{r,\mathrm{rel}},\label{eq: expected_valu_of_relative_velocities}\\
    \mathrm{d}v_{r,\mathrm{rel}} \equiv v_\mathrm{rel}P_{r,\mathrm{rel}}(v_\mathrm{rel})\mathrm{d}v_\mathrm{rel}.
\end{gather}
The small element of $\mathrm{d}v_{r,\mathrm{rel}}$ means an expected relative velocity between two DMSHs within $(v_{\mathrm{rel}},v_{\mathrm{rel}}+\mathrm{d}v_{\mathrm{rel}})$ at the distance $r$ from the centre of a host halo.

The next step is to calculate the collision frequency between two DMSHs of the same masses moving within the radius of the host halo. 
We define a parameter $\eta$ as the ratio of the virial radius of the host to the virial radius of a DMSH,
\begin{gather}
    \eta = \frac{R_\mathrm{200,\,sub}}{R_\mathrm{200,\,host}}=\left(\frac{M_\mathrm{sub}}{M_\mathrm{host}}\right)^{1/3},
\end{gather}
where $M_\mathrm{sub}$ is a mass of a colliding DMSH.
When the collisions of two DMSHs with the cross section $\sigma=\pi r_\mathrm{s,sub}^2 = \pi\eta^2R_\mathrm{200,host}^2/c_\mathrm{sub}^2$ moving the relative velocity $\mathrm{d}v_{r,\mathrm{rel}}$ inside the volume element $\mathrm{d} V=\mathrm{d} L^3$, the probability of one collision per area $\mathrm{d} L^2$ is $\sigma/\mathrm{d} L^2$. 
In the rest frame of one DMSH, another DMSH makes $\mathrm{d}v_{r,\mathrm{rel}}\mathrm{d} t/(2\mathrm{d} L)$ round trips of distance $\mathrm{d}L$ during a time $\mathrm{d}t$, thus the number of collisions of two DMSHs in a volume $\mathrm{d}V$ can be expressed by
\begin{gather}
    \frac{\sigma}{\mathrm{d} L^2}\cdot\frac{\mathrm{d}v_{r,\mathrm{rel}}\mathrm{d} t}{2\mathrm{d} L}=\frac{\sigma \mathrm{d}v_{r,\mathrm{rel}}\mathrm{d} t}{2\mathrm{d} V},
\end{gather}
where $\mathrm{d}V=4\pi r^2\mathrm{d}r$ for a spherical host halo.
We assume that the distribution of the number density of DMSHs in the host halo is assumed to be described by the NFW function,
\begin{gather}
    n(r)=N\nu(r)=
    \frac{Ng(c_\mathrm{host})}{4\pi R_\mathrm{200,\,host}^3}\frac{1}{x(1+x)^2},
\end{gather}
where $N$ is the total number of DMSHs in the host halo.
From the above, the number of collisions in a volume $\mathrm{d}V$ during a time $\mathrm{d}t$ is expressed as
\begin{align}
    \mathrm{d} k &= \frac{\sigma \mathrm{d}v_{r,\mathrm{rel}}\mathrm{d} t}{2\mathrm{d} V}\cdot \left(n\mathrm{d} V\right)^2\nonumber,\\
     &=\frac{N^2\eta^2 g(c_\mathrm{host})^2}{8R_\mathrm{200,host}^2c_\mathrm{sub}^2c_\mathrm{host}^2}\frac{ v_\mathrm{rel}P_{r,\mathrm{rel}}}{(1+x)^4}\mathrm{d} v_\mathrm{rel}\,\mathrm{d} t\,\mathrm{d} r. 
\end{align}

The collision frequency, which depends on the distance from the centre of the host halo and the relative velocity of the DMSHs, is written by
\begin{gather}
    \frac{c_\mathrm{sub}^2}{N^2\eta^2}\frac{\mathrm{d} k}{\,\mathrm{d}t \,\mathrm{d}r \,\mathrm{d}v_\mathrm{rel}}
    = \frac{g(c_\mathrm{host})^2}{8R_\mathrm{200,host}^2    c_\mathrm{host}^2}\frac{ v_\mathrm{rel}P_{r,\mathrm{rel}}}{(1+x)^4},\label{eq: Collision}
\end{gather}
where it is divided by free parameters of the colliding DMSH.
\begin{figure*}
    \centering
    \includegraphics[width=\textwidth]{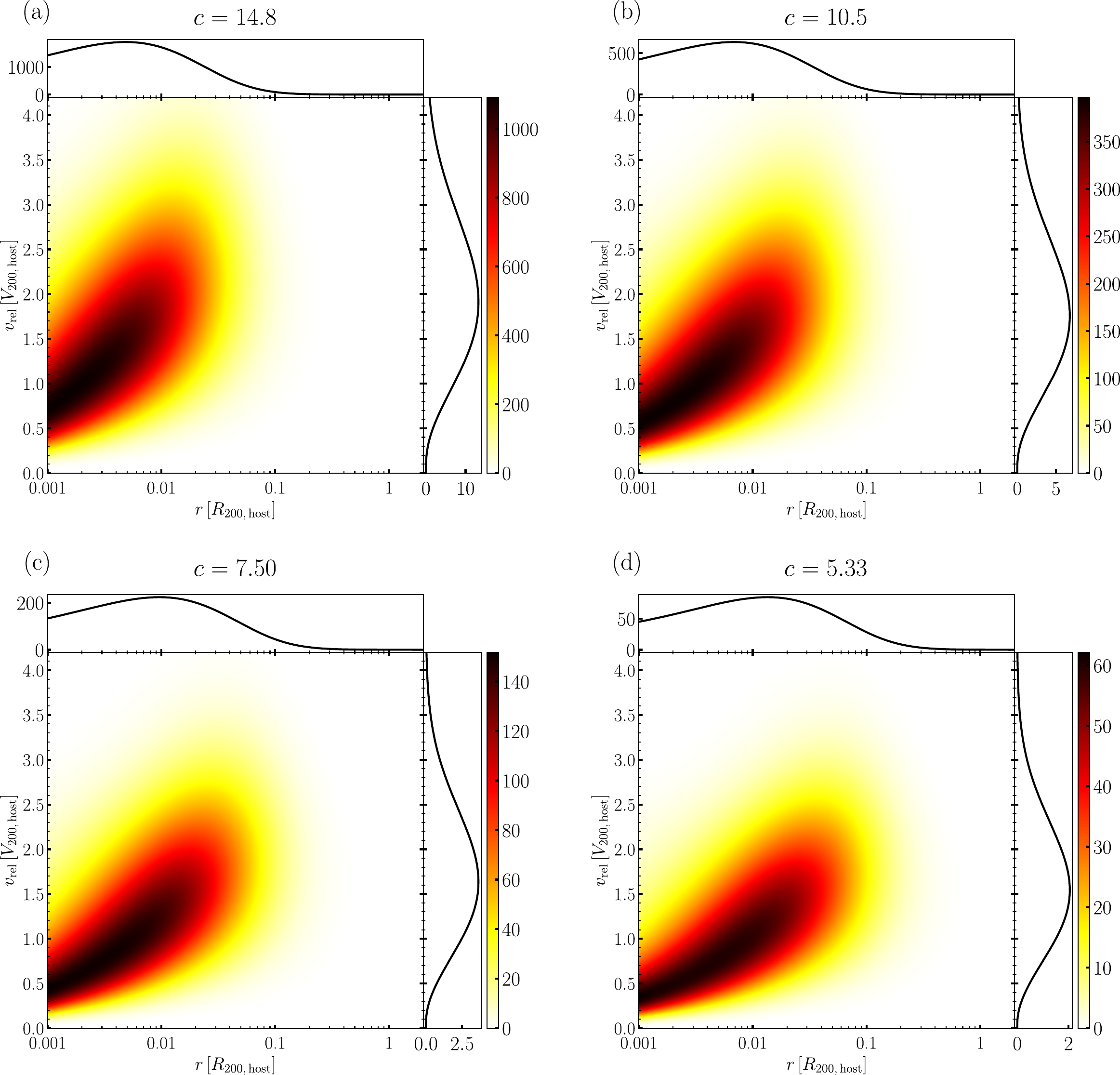}
    \caption{Distributions of collision frequency of {equation} \eqref{eq: Collision} between DMSHs within a host galaxy for (a) $c=14.8$, (b) $c=10.5$, (c) $c=7.50$, and (d) $c=5.33$, respectively. Top sub-panel in each panel: dependence of collision frequency on the radius of a host halo. Right sub-panel in each panel: dependence of collision frequency on the relative velocity between two DMSHs.}
    \label{fig: CollisionFreq}
\end{figure*}

The calculation result of equation \eqref{eq: Collision} is shown in Fig. \ref{fig: CollisionFreq}. 
The four panels correspond to the property of the host halo, (a) $c = 14.8$ for $M_\mathrm{host}=10^{8}\,\mathrm{M_\odot}$, (b) $c = 10.5$ for $M_\mathrm{host}=10^{10}\,\mathrm{M_\odot}$, (c) $c = 7.5$ for $M_\mathrm{host}=10^{12}\,\mathrm{M_\odot}$, and  (d) $c = 5.33$ for $M_\mathrm{host}=10^{14}\,\mathrm{M_\odot}$, respectively, using the $c\text{--}M$ relation \citep{PradaKlypinCuestaBetancort-RijoEtAl_2012_HaloConcentrationsStandard_MonNotRAstronSoc}. Table \ref{tab: col_freq} lists also the properties of host haloes.

The colours in the diagram correspond to the magnitude of the collision frequency, $c_\mathrm{sub}^2\mathrm{d}k/(N^2\eta^2\mathrm{d}t\mathrm{d}{r}\mathrm{d}v_\mathrm{rel})$. The horizontal axis represents the distance from the host centre, normalised by $R_{200}$. The vertical axis corresponds to the relative velocity between DMSHs, normalised by the circular velocity $V_{200}$.
The upper sub-panel shows the dependence of collision frequency on 
the radius integrated by the relative velocity, $c_\mathrm{sub}^2\mathrm{d}k/(N^2\eta^2\mathrm{d}t\mathrm{d}{r})$. 
The right sub-panel shows the dependence of collision frequency on the relative velocity between two DMSHs integrated by the radius, $c_\mathrm{sub}^2\mathrm{d}k/(N^2\eta^2\mathrm{d}t\mathrm{d}{v_\mathrm{rel}})$.
Depending upon the concentration $c$, the peak position and the peak relative velocity, which have the maximum collision probability $c_\mathrm{sub}^2\mathrm{d}k/(N^2\eta^2\mathrm{d}t\mathrm{d}{r})$ and $c_\mathrm{sub}^2\mathrm{d}k/(N^2\eta^2\mathrm{d}t\mathrm{d}{v_\mathrm{rel}})$, respectively, are fitted numerically by
\begin{align}
    \frac{r_\mathrm{peak}}{R_\mathrm{200,\,host}} &= (9.58\pm0.01)\times10^{-3}\left(\frac{c}{7.5}\right)^{-1.00\pm0.00},\\
    \frac{v_\mathrm{rel,\,peak}}{V_\mathrm{200,\,host}} &= (0.384\pm0.011)\left(\frac{c}{7.5}\right)^{0.766\pm0.015}\nonumber\\
    &\quad\quad+(1.26\pm0.01),
\end{align}
for $2\leq c\leq 30$, respectively.

In addition, we calculate the average relative velocity within the host halo as
\begin{align}
    \left<v_\mathrm{rel}\right> = \frac1A\int v_\mathrm{rel} \left(\frac{{c_\mathrm{sub}^2}}{N^2\eta^2}\frac{\mathrm{d} k}{\mathrm{d}v_\mathrm{rel}}\right)\mathrm{d}v_\mathrm{rel},
\end{align}
where $A$ is a normalisation constant given by 
\begin{align}
    A =  \frac{c_\mathrm{sub}^2}{N^2\eta^2}\int \left(\frac{\mathrm{d} k}{\,\mathrm{d}t \,\mathrm{d}r \,\mathrm{d}v_\mathrm{rel}}\right) \mathrm{d}t \,\mathrm{d}r \,\mathrm{d}v_\mathrm{rel}.
\end{align}
It should be noted that the velocity $\left<v_\mathrm{rel}\right>$ is different from the velocity $E[v_{r,\mathrm{rel}}]$ given by equation \eqref{eq: expected_valu_of_relative_velocities}, because $\left<v_\mathrm{rel}\right>$ is integrated over the whole region of a host halo.
As a result, the average relative velocity of the colliding DMSHs is derived by
\begin{align}
    \frac{\left<v_\mathrm{rel}\right>}{V_\mathrm{200,\,host}} &= (0.474\pm0.018)\left(\frac{c}{7.5}\right)^{0.0694\pm0.0184}\nonumber\\
    &\quad\quad+(1.34\pm0.02),
\end{align}
for $2\leq c\leq 30$.

\begin{figure}
    \centering
    \includegraphics[width=\columnwidth]{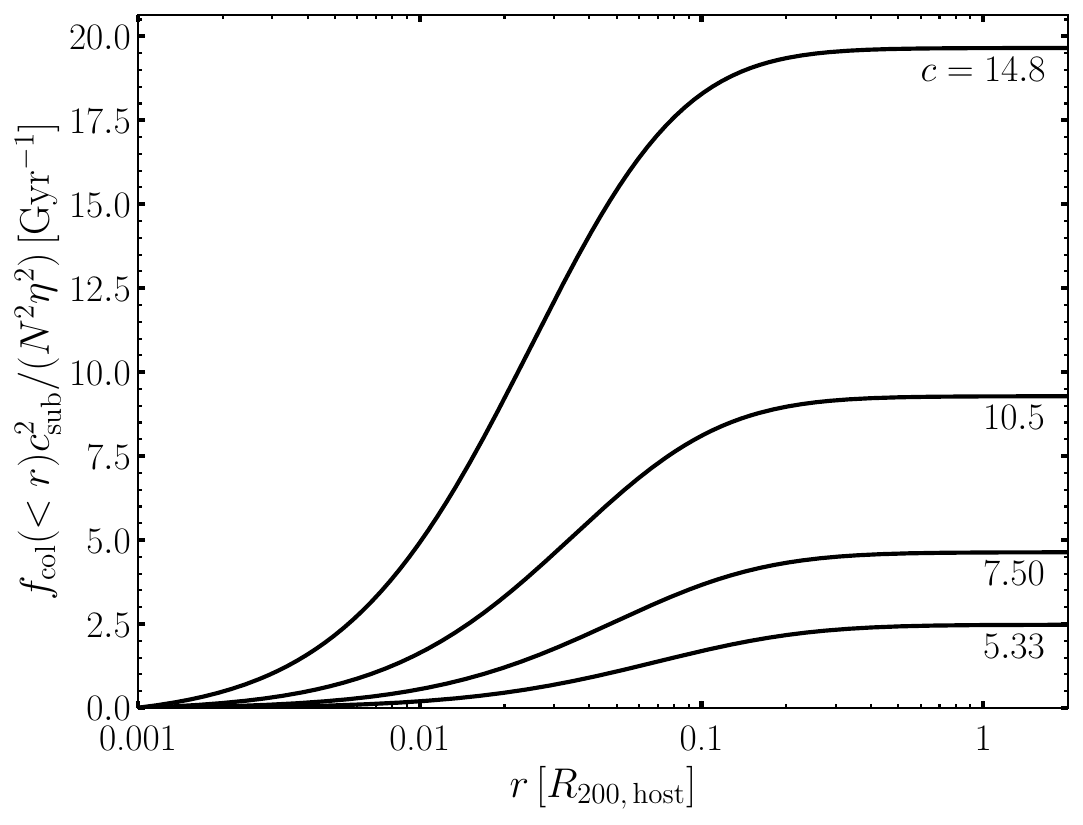}
    \caption{Radial profiles of cumulative collision frequency within $r$ divided by parameters $N^2 \eta^2 /c_\mathrm{sub}^2$, which are defined as equation \eqref{eq: CumColFreq} for $c=14.8,\,10.5,\,7.50,\text{ and } 5.33$, respectively.}
    \label{fig: CumColFreq}
\end{figure}

Fig. \ref{fig: CumColFreq} shows the cumulative collision frequency within $r$ for each concentration of the host halo, which is defined as

\begin{align}
    f_\mathrm{col}(<r) = \frac{N^2\eta^2}{c_\mathrm{sub}^2} \int_0^r \left(\frac{\mathrm{d}k}{\mathrm{d}t\,\mathrm{d}r'}\right)\mathrm{d}r'. \label{eq: CumColFreq}
\end{align}
{Fig. \ref{fig: CumColFreq} indicates most of the collisions occur within $\sim0.1 \,R_{200,\,\mathrm{host}}$. DMSHs might be tidally disrupted by the gravity of the host halo in the inner region of the peak position $r_\mathrm{peak}\sim 0.01 \,R_{200,\,\mathrm{host}}$. Here, we define the collision frequency in the outer region and the total collision frequency within the host halo as}
\begin{gather}
    f_\mathrm{col,\,0.01} \equiv \frac{N^2\eta^2}{c_\mathrm{sub}^2} \int_{0.01\,R_{200\,\mathrm{host}}}^{R_{200\,\mathrm{host}}} \left(\frac{\mathrm{d}k}{\mathrm{d}t\,\mathrm{d}r'}\right)\mathrm{d}r',\\
    f_\mathrm{col,\,tot} \equiv \frac{\mathrm{d}k}{\mathrm{d}t} =  \frac{N^2\eta^2}{c_\mathrm{sub}^2} \int_0^{R_{200\,\mathrm{host}} }\left(\frac{\mathrm{d}k}{\mathrm{d}t\,\mathrm{d}r'}\right)\mathrm{d}r',
\end{gather}
respectively.
In the case that DMSHs with $M_\mathrm{sub}=10^9\,\mathrm{M_\odot}$ move within the host galaxy of $M_\mathrm{host}=10^{12}\,\mathrm{M_\odot}$, the collision frequency at the outer region and the total collision frequency are given by
\begin{gather}
    f_\mathrm{col,\,0.01} =65.2\left(\frac{N}{500}\right)^2\,\mathrm{Gyr^{-1}},\\
    f_\mathrm{col,\,{tot}} = 74.2\left(\frac{N}{500}\right)^2\,\mathrm{Gyr^{-1}},
\end{gather}
respectively. The collision timescales corresponding to $f_\mathrm{col,\,0.01}$ and $f_\mathrm{col,\,tot}$ are equivalently given by
\begin{gather}
    \tau_{\mathrm{col},\,0.01} = 15.3\left(\frac{N}{500}\right)^{-2}\,\mathrm{Myr},\\
    \tau_\mathrm{col,\,tot} = 13.5\left(\frac{N}{500}\right)^{-2}\,\mathrm{Myr},
\end{gather}
respectively.

The subhalo collision frequency moving within a host halo with an NFW potential has not hitherto been quantitatively analysed.
The result implies that DMSHs orbiting the host halo experience frequent mutual collisions.
Therefore, it becomes clear that the galaxy collision model has a potential capability for the formation of dark-matter-deficient galaxies. Furthermore, a large number of traces of interactions between dwarf galaxies
observed in the local universe 
\citep[e.g., ][]{StierwaltBeslaPattonJohnsonEtAl_2015_TiNyTITANSROLE_ApJ, PaudelSmithYoonCalderon-CastilloEtAl_2018_CatalogMergingDwarf_ApJS, PoulainMarleauHabasDucEtAl_2022_HIObservationsMATLAS_A&A,ChhatkuliPaudelBachchanAryalEtAl_2023_FormingBlueCompact_MonthlyNoticesoftheRoyalAstronomicalSociety}
is a natural consequence compared to our results.
The future observation of more faint structures of subhalo collisions would provide the number and frequency of those collisions.

\begin{table*}
\centering
\caption{Properties of host halos: halo mass $M_{200}$, virial radius $R_{200}$, scale radius $r_\mathrm{s}$, circular velocity $V_{200}$, concentration $c$, {collision frequency for outer region of the 0.01 virial radius $f_{\mathrm{col},0.01}c_\mathrm{sub}^2/(N^2\eta^2)$, } and {total} collision frequency normalised by properties of colliding subhalos $f_\mathrm{col,\,{tot}}c_\mathrm{sub}^2/(N^2\eta^2)$.}
\label{tab: col_freq}
\begin{tabular}{ccccccc}
\toprule
$M_{200}\,[\mathrm{M_\odot}]$ &
  $R_{200}\,[\mathrm{kpc}]$ &
  $r_\mathrm{s}\,[\mathrm{kpc}]$ &
  $V_{200}\,[\mathrm{km\,s^{-1}}]$ &
  $c$ &
  {$f_{\mathrm{col}, 0.01}c_\mathrm{sub}^2/(N^2\eta^2)\,[\mathrm{Gyr^{-1}}]$} &
  $f_\mathrm{col,\,{tot} }c_\mathrm{sub}^2/(N^2\eta^2)\,[\mathrm{Gyr^{-1}}]$  \\ \toprule
$10^{8}$  & $9.82$ & $0.063$ & $6.62$ & $14.8$ & {$14.7$} & $19.7$ \\
$10^{10}$ & $45.6$ & $4.32$  & $30.7$ & $10.5$ & {$7.63$} & $9.28$ \\
$10^{12}$ & $212$  & $282.$  & $143$  & $7.50$ & {$4.07$} & $4.63$ \\
$10^{14}$ & $982$  & $184$   & $662$  & $5.33$ & {$2.28$} & $2.48$ \\ \bottomrule
\end{tabular}
\end{table*}

\section{Analytical Model} 
We consider collisions between DMSHs in the dark matter halo of the massive host galaxy. 
DMSHs are bound and moving within the dark matter potential of massive galaxies and can collide at various velocities.
This section evaluates the physical processes in a model with head-on collisions of DMSHs, assuming that a DMSH is composed of a mass ratio of the dark matter to the gas component of 5.36. 
$M_\mathrm{dm}$ and $M_\mathrm{gas}$ denote the masses of gas and dark matter in the DMSH, respectively. 

\subsection{subhaloes merger}
In the case of slow collisions that are sufficiently slower than the escape velocity of the system, DMSHs merge into one self-gravitationally bound system after the collision. In the deep gravitational potential created by the dark matter, the gas will self-gravitationally contract, eventually leading to the formation of a dark-matter-dominated galaxy.

We assume that the gravitational potential of a DMSH with mass $M_\mathrm{sub}$ may be approximated by an NFW potential. Here, the physical quantities of a host halo in the equation \eqref{eq: NFWpot} are replaced by those of DMSHs.
If two DMSHs that consist of dark matter collide head-on with a relative velocity $v_\mathrm{rel}$, they have the energy condition,
\begin{gather}
        \frac12 v_\mathrm{rel}^2 \leq -\Phi_\mathrm{NFW}(R_\mathrm{200, merged}), \\
        R_\mathrm{200,merged}=\left(\frac{3\cdot 2M_\mathrm{sub}}{4\pi \rho_{200}}\right)^{1/3}, \label{eq: merger}
\end{gather}
in order to merge. 
Therefore, the critical relative velocity is given by
\begin{align}
    v_\mathrm{merger} &=2M_\mathrm{sub}^{1/3}\left(\frac{2\pi \rho_{200}}{3}\right)^{1/6}\nonumber\\
    &\quad\quad \times \left(\frac{G\ln{(1+c_\mathrm{sub})}}{\ln{(1+c_\mathrm{sub})}-c_\mathrm{sub}/(1+c_\mathrm{sub})}\right)^{1/2},\label{eq: v_merger0}\\
    &\simeq 14.5\left(\frac{M_\mathrm{sub}}{10^9\,\mathrm{M_\odot}}\right)^{0.34} \,\mathrm{km\,s^{-1}}, \label{eq: v_merger1}
\end{align}
The function of concentration $c(M)$ can be approximated by a simple power-law using $c\text{--}M$ relation introduced by \citet{PradaKlypinCuestaBetancort-RijoEtAl_2012_HaloConcentrationsStandard_MonNotRAstronSoc}.

\subsection{Shock-breakout} \label{sec: ShockBreakout}

In the case of high-speed collisions such as $v_\mathrm{rel}\gg v_\mathrm{merger}$, a shock wave is formed and propagates in the gaseous medium. Then no galaxies form because shock-breakout ejects most of the gas from the system. To estimate the propagation of a strong adiabatic shock through an ideal gas, we consider a simple one-dimensional model of the two colliding gas clouds.
We assume that two clouds have mass $M_\mathrm{gas}$, radius $r_\mathrm{s}$, uniform density $\rho_0=3M_\mathrm{gas}/4\pi r_\mathrm{s}^3$ and specific internal energy $u_0=0$.
The cloud centres are initially at $\pm r_\mathrm{s}$, and the bulk velocities of the clouds are $\mp v_\mathrm{rel}/2$, respectively.
The subscript 0 indicates the physical quantities before the collision of the DMSHs, and the subscript 1 indicates the physical quantities after the shock wave generated by the collision has passed through.

According to the Rankine--Hugoniot (RH) condition, the density of the shocked clouds for strong adiabatic shocks is given by

\begin{align}
    \rho_1 = \frac{\gamma+1}{\gamma-1}\rho_0,
\end{align}
where $\gamma$ is the specific heat ratio. We have used $\gamma=5/3$.
We assume that the kinetic energy of the system is roughly converted into thermal energy as $u_1 = (v_\mathrm{rel}/2)^2/2$ and $v_1 = 0$ for the energy conservation.
Using the RH condition for the mass conservation, we derive the shock velocity propagating within each cloud with an initial position at $\pm r_\mathrm{s}$ as
\begin{align}
    v_\mathrm{shock} = \pm\frac{\gamma-1}{2}\frac{v_\mathrm{rel}}{2},
\end{align}
respectively.

When the shock waves reach their cloud surface, most of the gas is ejected from the system. 
The shock-crossing time is defined by 
\begin{align}
 t_\mathrm{cross}=\frac{2r_\mathrm{s}}{v_\mathrm{rel}/2+v_\mathrm{shock}}. \label{eq: ShockTime}
\end{align}
We compare it with other timescales, such as the cooling time or the free-fall time to estimate the critical relative velocity of shock-breakout.
The cooling time in the shocked gaseous medium is 
\begin{gather}
 t_\mathrm{cool}=\frac{k_\mathrm{B}m_\mathrm{p}\mu_1T_1}{(\gamma-1)\rho_1\Lambda(T_1,Z)},\quad T_1 = \frac{(\gamma-1)\mu_1 m_\mathrm{p}u_1}{k_\mathrm{B}}, \label{eq: CoolingTime}
\end{gather}
where $k_\mathrm{B}$ is the Boltzmann constant, $m_\mathrm{p}$ is the proton mass, $T$ is the temperature, $\mu$ is the mean molecular weight, and 
$\Lambda(T, Z)$ is the cooling function as a function of the temperature and metallicity $Z$.

Here, we assume $\Lambda(T, 0.1\,\mathrm{Z_\odot})$ 
in collisional ionization equilibrium (CIE) given by \texttt{MAPPINGS V} \citep{SutherlandDopita_2017_EffectsPreionizationRadiative_ApJS,SutherlandDopitaBinetteGroves_2018_MAPPINGSAstrophysicalPlasma_AstrophysicsSourceCodeLibrary}.

It is obvious that effective radiative cooling, $t_\mathrm{cool}< t_\mathrm{cross}$, promotes and enhances the formation of galaxies, while the shock-breakout, $t_\mathrm{cross}< t_\mathrm{cool}$, prohibits or suppresses the formation of galaxies.
The critical relative velocity of $t_\mathrm{cool} = t_\mathrm{cross}$ is fitted by 
\begin{gather}
    v_\mathrm{crit} \simeq 691\left(\frac{M_\mathrm{sub}}{10^9\,\mathrm{M_\odot}}\right)^{0.06} \,\mathrm{km\,s^{-1}}. \label{eq: v_SB} 
\end{gather}

\begin{figure}
    \centering
    \includegraphics[width=\columnwidth]{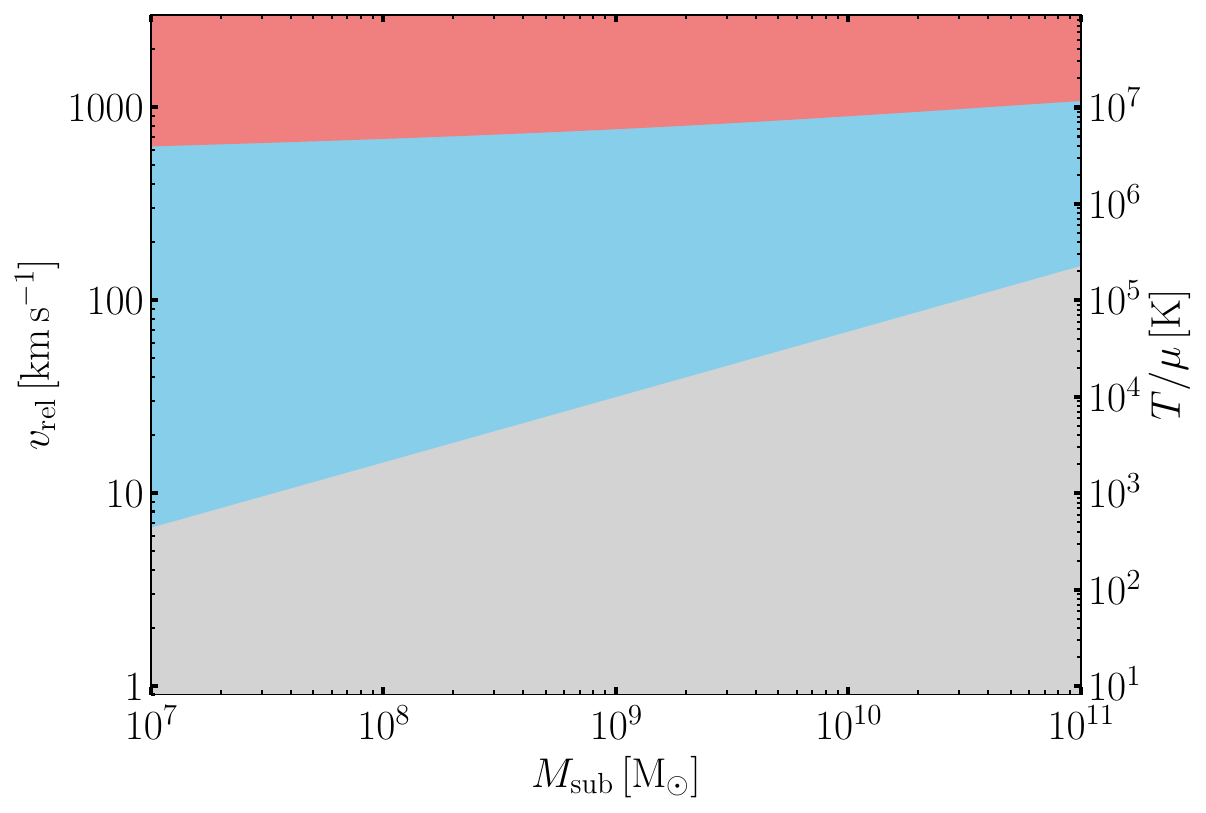}
    \caption{Results of analytical models assuming {0.1} solar metallicity. Grey region: velocity condition satisfied with the formation of the dark-matter-dominated galaxies. Blue region: velocity condition satisfied with the formation of dark-matter-deficient galaxies. Red region: no galaxy form to occur shock-breakout. {The right axis indicates the temperature divided by the mean molecular weight $T/\mu$, which corresponds to the kinetic energy of the {relative velocity $v_\mathrm{rel}$}.}}
    \label{fig: 1Dresult}
\end{figure}
Fig. \ref{fig: 1Dresult} shows the results of the analytical models. The grey region is the velocity condition satisfied with the formation of the dark-matter-dominated galaxies. The blue region is the velocity condition satisfied with the formation of dark-matter-deficient galaxies. The red region is no galaxy form to occur shock-breakout. We derived the critical {relative} velocities for the bifurcation sequence of the formation of dark-matter-dominated galaxies and dark-matter-deficient galaxies.

\section{Numerical Model}

A simple analytical model was used in the previous section to provide physical insight into DMSH collisions and galaxy formation. However, this analytical model contains several assumptions that need to be validated. We, therefore, perform a realistic $N$-body/hydrodynamic simulation of DMSH collisions, incorporating star formation and supernova feedback, to reveal the formation processes of dark-matter-deficient galaxies and dark-matter-dominated galaxies.

\subsection{Simulation set-up}

The simulation adopts the hierarchical tree algorithm for self-gravity and the three-dimensional smoothed particle hydrodynamics \citep[SPH: ][]{Lucy_1977_NumericalApproachTesting_TheAstronomicalJournal,GingoldMonaghan_1977_SmoothedParticleHydrodynamics_MonNotRAstronSoc} method for gas dynamics.

The acceleration of the one particle at a position $\bm{r}_i$ in a gravitational field consisting of $N$ particles such as dark matter, star and gas particles is obtained as
\begin{gather}
    \frac{\mathrm{d}\bm{v}_i}{\mathrm{d}t}=-\sum_{j=1}^N\frac{G m_j (\bm{r}_i-\bm{r}_j)}{(r_{ij}^2 + \epsilon^2)^{3/2}},
\end{gather}
where $G$ is the gravitational constant, $\bm{v}_j$ and $m_j$ are the velocity vector and the mass of a particle at $\bm{r}_j$, respectively, and $r_{ij}=|\bm{r}_i-\bm{r}_j|$. The softening length $\epsilon$ is a free parameter introduced to avoid numerical divergence.


In the SPH formulation, the gas density of one particle at a position $\bm{r}_i$ is given by
\begin{gather}
    \rho_i = \sum^{N_\mathrm{neigh}}_{j=1} m_jW(r_{ij},h_i), \label{eq: SPHdens}
\end{gather}
where $W(r,h)$ is the smoothing kernel, $h$ is the smoothing length, and $N_\mathrm{neigh}=200$ is the number of neighbour particles. We adopt the \cite{Wendland_1995_PiecewisePolynomialPositive_AdvComputMath} $C^4$ function,
\begin{gather}
    W(r,h)=\frac{495}{32\pi h^3}\begin{cases}
    (1-q)^6(1+6q+\frac{35}{3}q^2),\quad & q\leq1,\\
    0,&q>1,
    \end{cases}
\end{gather}
where $q=r/h$, to avoid the clumping instability \citep{DehnenAly_2012_ImprovingConvergenceSmoothed_MonNotRAstronSoc,ZhuHernquistLi_2015_NUMERICALCONVERGENCESMOOTHED_ApJ}. 
We basically followed the formulation of SPH introduced by \cite{SpringelHernquist_2002_CosmologicalSmoothedParticle_MonthlyNoticesoftheRoyalAstronomicalSociety}.
The smoothing length $h_i$ of each particle is determined by 
\begin{gather}
    \frac{4\pi}{3}h_i^3\rho_i=\overline{m}N_\mathrm{neigh}, \label{eq: SPHsmth}
\end{gather}
where $\overline{m}$ is the average mass of gas particles. These equations \eqref{eq: SPHdens} and \eqref{eq: SPHsmth} need to be solved implicitly for $\rho_i$ and $h_i$. 
However, the minimum value of smoothing length $h$ is set to gravitational softening $\epsilon$ in order to match the spatial resolution.

The momentum equation for a gas particle is given by
\begin{align}
    \frac{\mathrm{d}\bm{v}_i}{\mathrm{d}t}=-\sum_j m_j\Big [&f_i\frac{p_i}{\rho_i^2}\nabla_iW(r_{ij},h_i)+f_j\frac{p_j}{\rho_j^2}\nabla_iW(r_{ij},h_j)\nonumber\\
    & +\Pi_{ij}\nabla_i\overline{W}_{ij}\Big]\label{eq: Motion}
\end{align}
where $p_i$ is the pressure of the gas particle, $f_i$ is defined by 
\begin{gather}
    f_i=\left(1+\frac{h_i}{3\rho_i}\frac{\partial\rho_i}{\partial h_i}\right)^{-1},
\end{gather}
and $\overline{W}_{ij}$ is a symmetrised kernel, 
\begin{align}
    \overline{W}_{ij} = \frac12 \left[W(r_{ij},h_i)+W(r_{ij},h_j)\right].
\end{align}

Artificial viscosity $\Pi_{ij}$ is necessary for the proper handling of shocks.
For numerical stability, the entropic function $A_i=p_i/\rho_i^\gamma$ rather than the specific internal energy $u_i$ is used to calculate the thermodynamic evolution of gas particles. The entropy equation is given by
\begin{align}
    \frac{\mathrm{d}A_i}{\mathrm{d}t}&=\frac12\frac{\gamma-1}{\rho_i^{\gamma-1}}\sum_j m_j\Pi_{ij}\bm{v}_{ij}\cdot\nabla_i\overline{W}_{ij}, \label{eq: Entropy}
\end{align}
where $A_i$ is conserved in adiabatic flow, but it is generated by artificial viscosity via shocks. 

\subsubsection{Artificial viscosity} \label{sec: AV}
In this paper, we adopt \citeauthor{Monaghan_1997_SPHRiemannSolvers_JournalofComputationalPhysics}'s \citeyearpar{Monaghan_1997_SPHRiemannSolvers_JournalofComputationalPhysics} artificial viscosity $\widetilde{\Pi}_{ij}$ combined with \citeauthor{Balsara_1995_NeumannStabilityAnalysis_JournalofComputationalPhysics}'s \citeyearpar{Balsara_1995_NeumannStabilityAnalysis_JournalofComputationalPhysics} switch $F_i$.
The artificial viscosity is expressed as 
\begin{gather}
    \Pi_{ij}=\frac{F_i+F_j}{2}\widetilde{\Pi}_{ij},
\end{gather}
where 
\begin{gather}
    \widetilde{\Pi}_{ij}=\begin{cases}-\alpha\cfrac{v_{ij}^\mathrm{sig}w_{ij}}{\rho_i+\rho_j}\quad &\bm{v}_{ij}\cdot\bm{r}_{ij}<0,\\
    0&\bm{v}_{ij}\cdot\bm{r}_{ij}\geq0,
    \end{cases}\\
    v_{ij}^\mathrm{sig}=c_{\mathrm{s},i}+c_{\mathrm{s},j}-3w_{ij},\\
    w_{ij}=\bm{v}_{ij}\cdot\bm{r}_{ij}/|\bm{r}_{ij}|,
\end{gather}
and 
\begin{gather}
      F_{i} = \frac{|\nabla_i\cdot\bm{v}_i|}{|\nabla\cdot\bm{v}_i|+|\nabla_i\times\bm{v}_i|+0.0001 c_{\mathrm{s},i}/h_i}.
\end{gather}
In order to handle strong shock waves generated by the galaxy collisions, we put the parameter $\alpha=5$, which adjusts the strength of the artificial viscosity.

\begin{figure*}
    \centering
    \includegraphics[width=\textwidth]{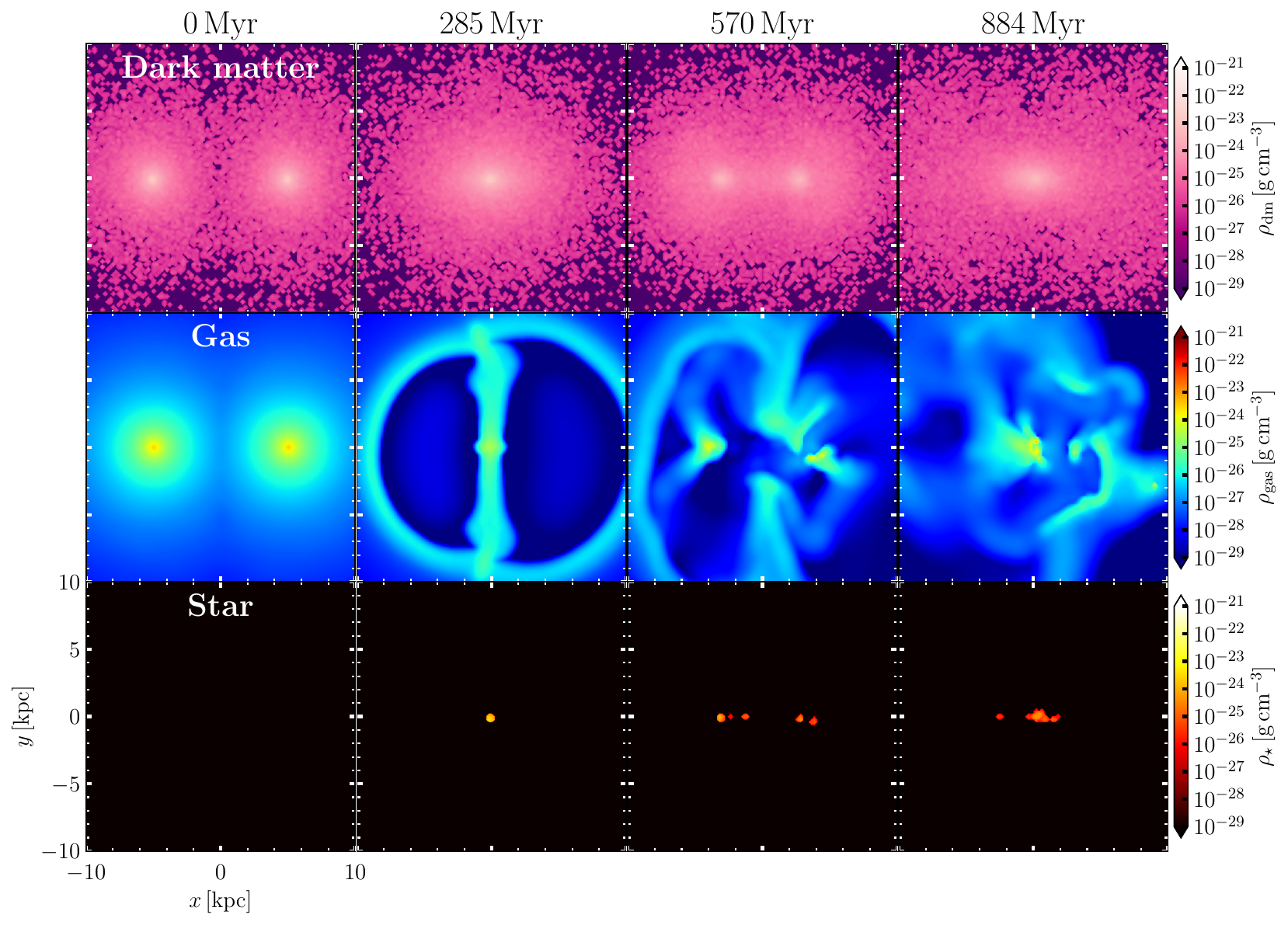}
    \caption{Snapshots of dark matter density (top), gas density (middle) and stellar density (bottom) of collision simulation between DMSHs with $10^9\,\mathrm{M_\odot}$ at {the relative velocity of $20\,\mathrm{km\,s^{-1}}$}. All the colour bars for mass density range from $10^{-29}$ to $10^{-21}\,\mathrm{g\,cm^{-3}}$ .
    From left to right, $t=0,\,285,\,570\text{ and }884\,\mathrm{Myr}$, respectively. A dark-matter-dominated galaxy forms in the case of this velocity. The masses of star, gas and dark matter enclosed within the bound radius $r_\mathrm{bound}=16.1\,\mathrm{kpc}$ are $M_\star = 5.19\times10^6\,\mathrm{M_\odot},\,M_\mathrm{gas}=2.88\times10^7\,\mathrm{M_\odot}\text{ and }M_\mathrm{DM}=1.16\times10^9\,\mathrm{M_\odot}$ at $t=4.7\,\mathrm{Gyr}$, respectively.}
    \label{fig: 1e9Msun_10kms}
\end{figure*}

\begin{figure*}
    \centering
    \includegraphics[width=\textwidth]{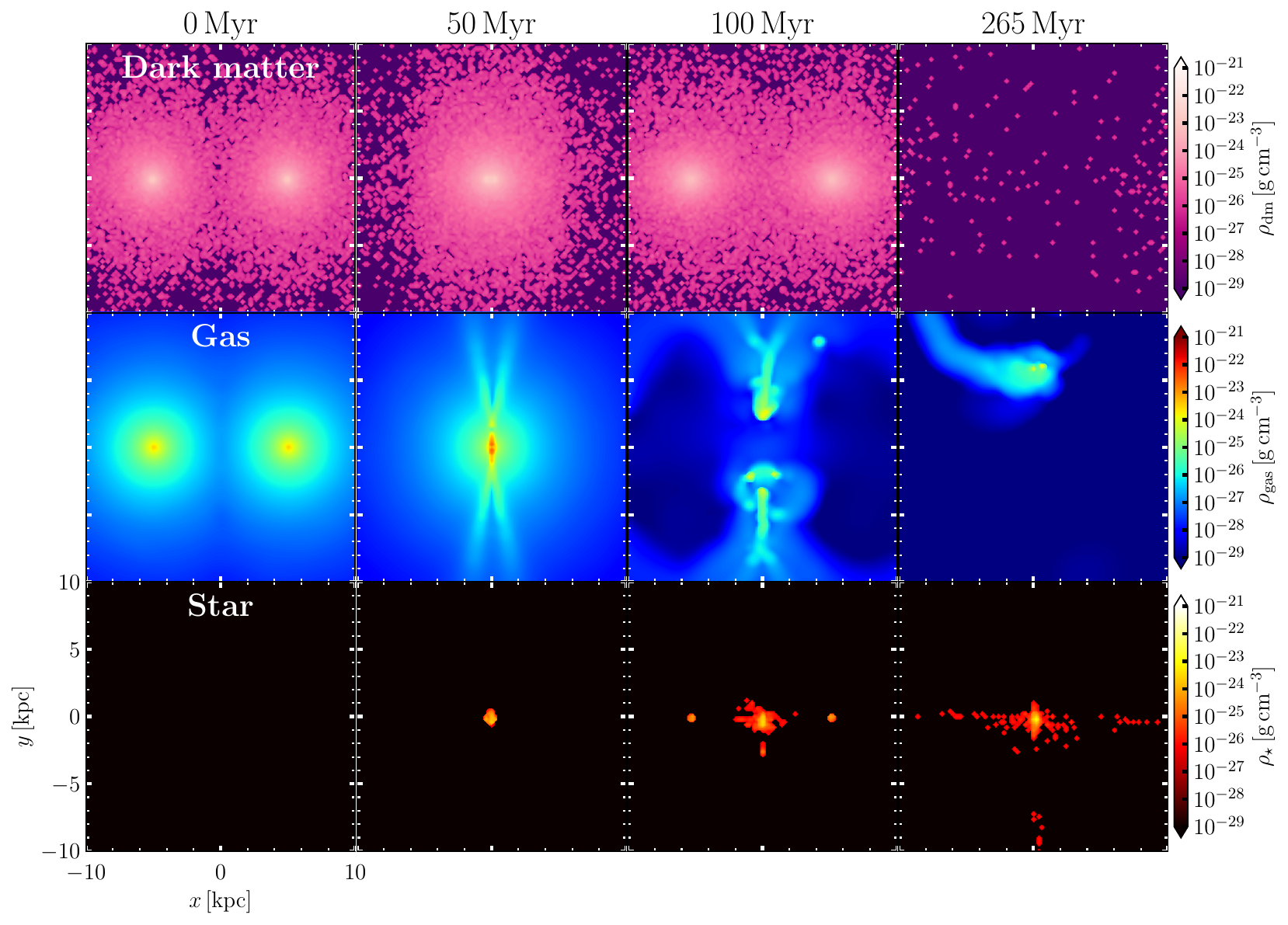}
    \caption{Same as Figure \ref{fig: 1e9Msun_10kms}, but {the relative velocity is $200\,\mathrm{km\,s^{-1}}$}. From left to right, $t=0,\,50,\,100,\text{ and }265\,\mathrm{Myr}$, respectively. A dark-matter-deficient galaxy forms in the case of this velocity.
    The masses of star, gas and dark matter are $M_\star = 1.34\times10^7\,\mathrm{M_\odot},\,M_\mathrm{gas}=4.02\times10^5\,\mathrm{M_\odot}\text{ and }M_\mathrm{DM}=0\,\mathrm{M_\odot}$ at $t=3\,\mathrm{Gyr}$, respectively.}
    \label{fig: 1e9Msun_100kms}
\end{figure*}

\begin{figure}
    \centering
    \includegraphics[width=\columnwidth]{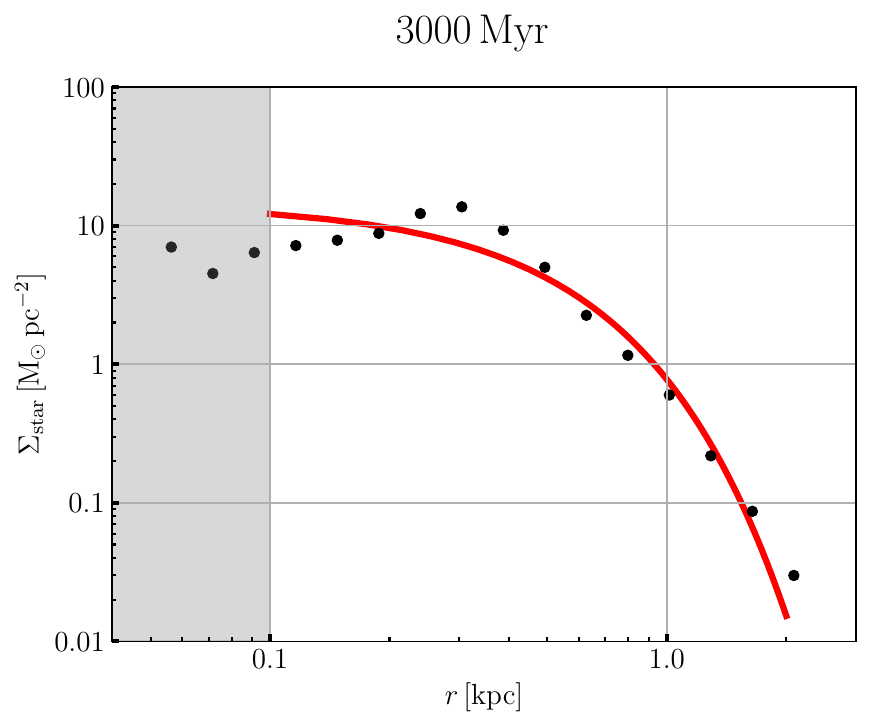}
    \caption{Radial profile of stellar surface density $\Sigma_\mathrm{star}$ in the face-on plane. The black points are stellar surface densities of a simulated galaxy averaged for each distance from the centre of mass for the collision simulation of {$200\,\mathrm{km\,s^{-1}}$}. The solid line represents the S\`{e}rsic curve of the effective radius $r_\mathrm{e}=0.5\,\mathrm{kpc}$ and S\`{e}rsic index $n_\text{S\`{e}rsic}=0.8$.
    {The region below the spatial resolution $\epsilon = 0.1\,\mathrm{kpc}$ is shaded in grey.}}
    \label{fig: ColDens}
\end{figure}

\begin{figure}
    \centering
    \includegraphics[width=\columnwidth]{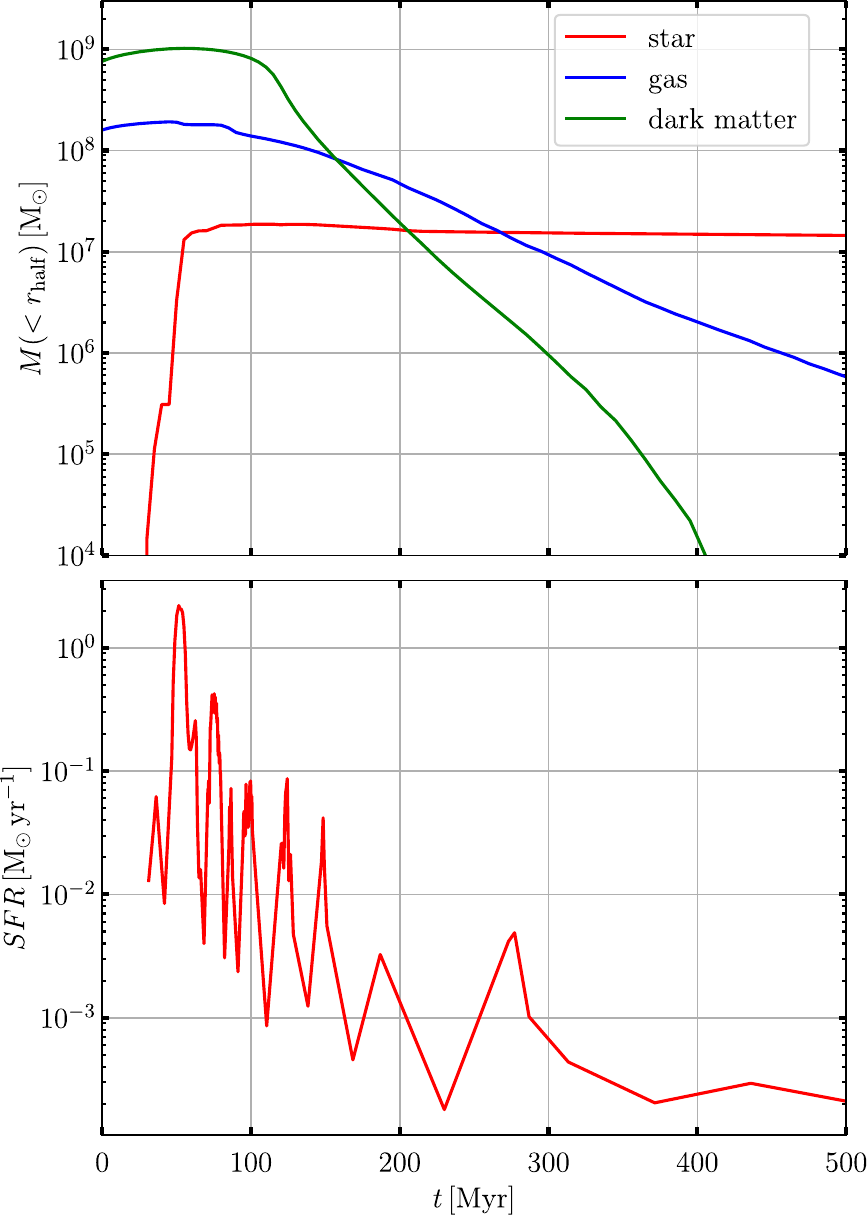}
    \caption{Evolution of the DMSH collision for the {relative velocity of $200\,\mathrm{km\,s^{-1}}$}. Top panel: evolution of the enclosed mass within a half mass radius of a DMSH in the initial condition. The red, blue and green lines are enclosed masses of star, gas and dark matter, respectively. 
    Bottom panel: the history of the overall star formation rate for the {relative velocity of $200\,\mathrm{km\,s^{-1}}$ }in the simulation box.}
    \label{fig: SFR_EnclosedMass}
\end{figure}

\begin{figure*}
    \centering
    \includegraphics[width=\textwidth]{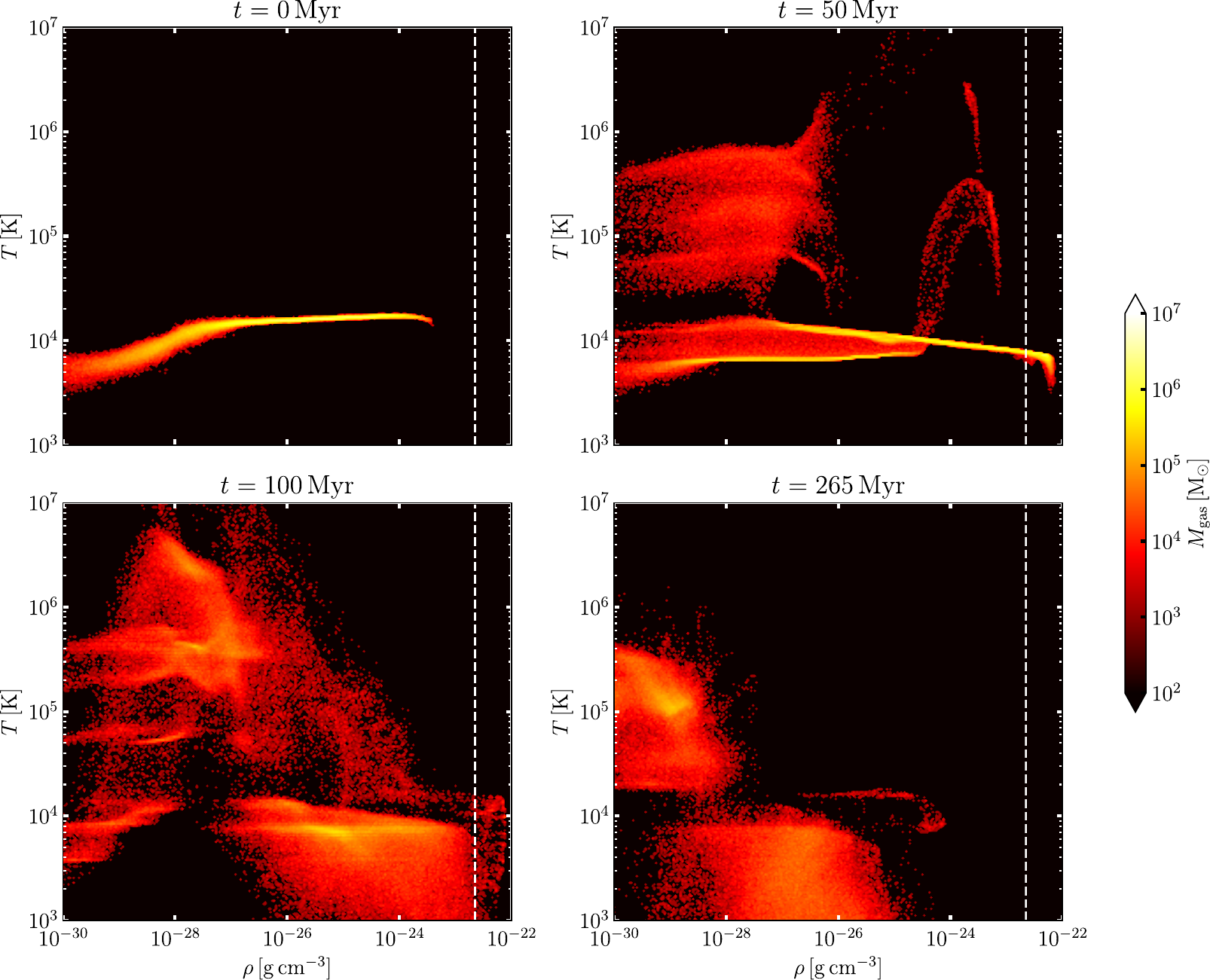}
    \caption{Density-temperature diagram for overall gas particles in the simulation box {for the relative velocity $200\,\mathrm{km\,s^{-1}}$}. Colour represents the mass of gas particles. The white dashed line is the star formation threshold in this simulation.}
    \label{fig: RhoT}
\end{figure*}

\begin{figure*}
    \centering
    \includegraphics[width=\textwidth]{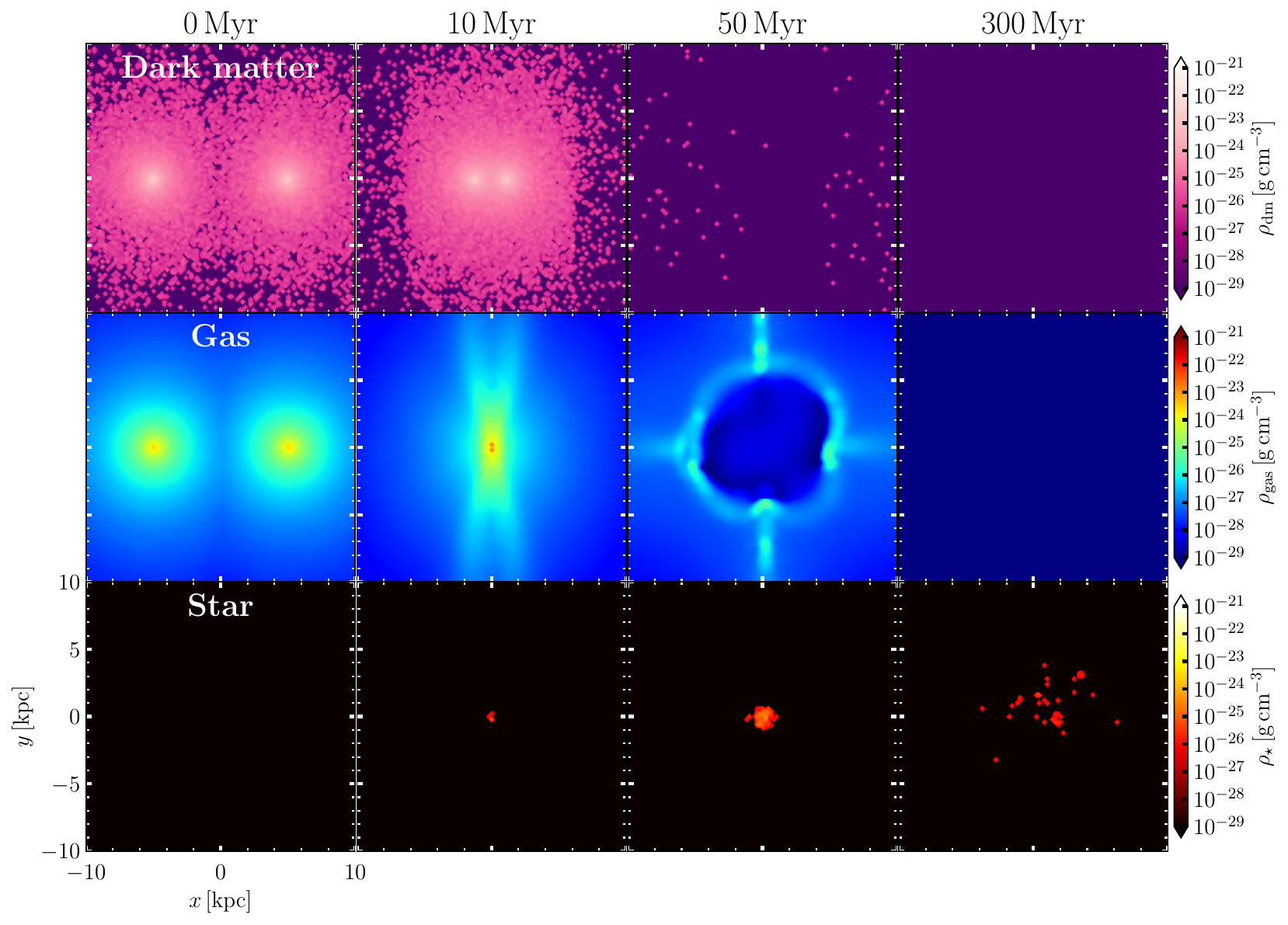}
    \caption{Same as Figure \ref{fig: 1e9Msun_10kms}, but relative velocity is $1200\,\mathrm{km\,s^{-1}}$. From left to right, $t=0,\,10,\,50,\text{ and }200\,\mathrm{Myr}$, respectively.}
    \label{fig: 1e9Msun_600kms}
\end{figure*}

\begin{figure}
    \centering
    \includegraphics[width=\columnwidth]{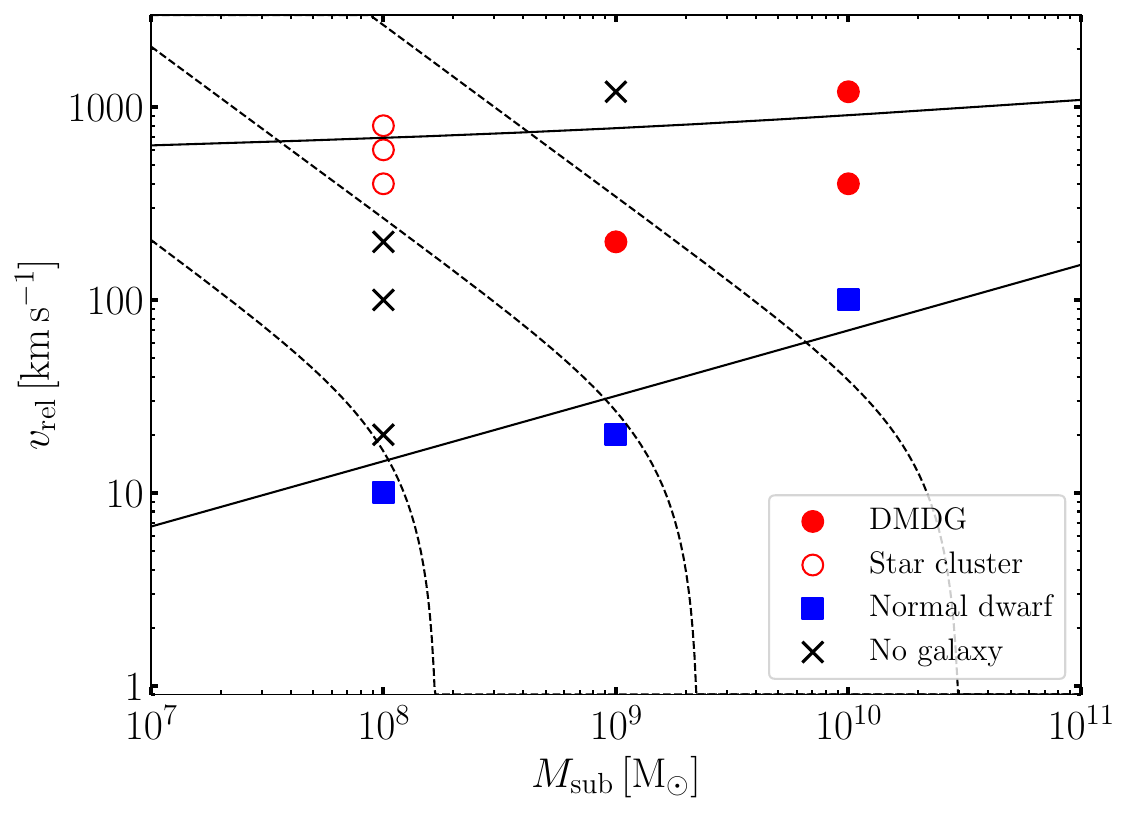}
    \caption{Results of simulations performed for $Z=0.1\,\mathrm{Z_\odot}$. The horizontal axis is the mass of a colliding DMSH, and the dark matter mass to gas mass ratio is $5.36$. The vertical axis is the {relative} velocity.
    {Red filled circles, red open circles, and blue squares indicate the formation of a dark-matter-deficient galaxy, star cluster, and a dark-matter-dominated galaxy, respectively.} The crosses indicate the result of no galaxy formation. The upper and lower solid lines are shock-breakout and merger conditions, respectively. The dashed lines are the Jeans criteria for the isothermal collisions of $10^4\,\mathrm{K}$. These dashed lines, from left to right, correspond to cases for the initial gas mass to Jeans mass ratio $\beta = 1.0,\,0.1,\,0.01$, respectively. }
    \label{fig: strong[Z]=-1}
\end{figure}

\subsubsection{Radiative cooling}
As radiative cooling plays a crucial role in galaxy formation and evolution, a term for energy dissipation due to radiative cooling needs to be added to the entropy equation:
\begin{align}
    \left(\frac{\mathrm{d}A}{\mathrm{d}t}\right)_\mathrm{cool}=\frac{\gamma-1}{\rho^\gamma}\left(\frac{\mathrm{d}u}{\mathrm{d}t}\right)_\mathrm{cool}
\end{align}
where
\begin{gather}
    \left(\frac{\mathrm{d}u}{\mathrm{d}t}\right)_\mathrm{cool}=-\frac{n^2 \Lambda(u,Z)}{\rho} = -\frac{\rho\Lambda(u,Z)}{\mu^2m_\mathrm{p}^2}. \label{eq: Cooling}
\end{gather}
Here, $n$, $\mu$ and $m_\mathrm{p}$ are the number density of the gas, the mean molecular weight and the proton mass, respectively.
$\Lambda$ is the cooling function of the specific internal energy $u$ and the metallicity $Z$. In order to solve the cooling equation \eqref{eq: Cooling}, we use the Exact Integration (EI) scheme \citep{Townsend_2009_EXACTINTEGRATIONSCHEME_ApJS}.
Integrating from time $t^n$ to $t^{n+1}=t^n+\Delta t$, the equation \eqref{eq: Cooling} becomes
\begin{gather}
    \int_{u_i^n}^{u_i^{n+1}}\frac{\mu(u)^2}{\Lambda(u,Z)}\mathrm{d}u = -\frac{\rho_i}{m_\mathrm{p}^2}\Delta t,
\end{gather}
where we also take into account the time evolution of the mean molecular weight.
Then, using the temporal evolution function which is defined by 
\begin{gather}
    Y(u)=\frac{\Lambda_\mathrm{ref}}{\mu_\mathrm{ref}^2u_\mathrm{ref}}\int_u^{u_\mathrm{ref}}\frac{\mu(u)^2}{\Lambda(u,Z)}\mathrm{d}u,
\end{gather}
the cooling equation \eqref{eq: Cooling} becomes
\begin{gather}
    u_i^{n+1}=Y^{-1}\left[Y(u_i^n)+\frac{\Lambda_\mathrm{ref}}{\mu_\mathrm{ref}^2u_\mathrm{ref}}\frac{\rho_i}{m_\mathrm{p}^{2}}\Delta t\right].\label{eq: EI_scheme}
\end{gather}
$Y(u)$ can be obtained as a table by fitting the cooling function $\Lambda$ with a piecewise power law. By using this integrated function $Y(u)$, the time evolution equation can be solved taking into account the temperature dependence of the cooling rate and is not sensitive to the size of the time step.

\subsubsection{Star formation}
We install the algorithms for the star formation and the resulting energetic “feedback” from young, massive stars. 
The simulation of the galaxy formation has usually an insufficient resolution to resolve these processes directly and instead adopts sub-grid physics tuned to match large-scale observational constraints. The star formation is realised by the conversion of the SPH particle into a stellar particle with an initial mass function (IMF).

In this study, we assume that a stellar particle has \cite{Salpeter_1955_LuminosityFunctionStellar_TheAstrophysicalJournal} IMF.
For a single SPH particle, it is converted to a stellar particle when the conditions are satisfied \citep{Katz_1992_DissipationalGalaxyFormation_ApJ}:  
\begin{gather}
    n_\mathrm{H}>10\,\mathrm{cm^{-3}} \label{eq: nSF}\\
    r\leq p = 1-\exp{\left(-\frac{C_\star \Delta t}{t_\mathrm{ff}}\right)}
\end{gather}
where $n_\mathrm{H}$ is the number density of hydrogen, $r$ is the random value between 0 and 1, $C_\star=1$ is the constant value corresponding to the star formation efficiency and $t_\mathrm{ff}$ is the local free-fall time.

\subsubsection{Supernova feedback}
When a star with a mass of more than $8\,\mathrm{M_\odot}$ reaches the end of its life, it undergoes a supernova explosion, releasing about $10^{51}\,\mathrm{erg}$ per star into the surrounding gas. This supernova feedback heats up the surrounding gas, leading to a decrease in the star formation rate of the galaxy. Assuming a \citet{Salpeter_1955_LuminosityFunctionStellar_TheAstrophysicalJournal} IMF with the upper mass of $60\,\mathrm{M_\odot}$ and the lower mass of $0.1\,\mathrm{M_\odot}$
, the number of massive stars $(>8\,\mathrm{M_\odot})$ in a stellar particle of mass $m_\star$ is
\begin{gather}
    N_\mathrm{SN}\simeq 73.1\left(\frac{m_\star}{10^4\,\mathrm{M_\odot}}\right).
\end{gather}
Of all the stars that undergo supernova explosions, the shortest lifetime is $5.4 \,\mathrm{Myr} \,(60\,\mathrm{M_\odot})$ and the longest is $43 \,\mathrm{Myr} \,(8 \,\mathrm{M_\odot})$. During this period, the feedback energy is released from the star to the gas.
If we assume that the energy is equally distributed to neighbour SPH particles of the star particle, the energy received by the SPH particle per time step $\Delta t$ is
\begin{gather}
    \Delta E = \frac{L_\mathrm{SN}N_\mathrm{SN}\Delta t }{N_\mathrm{neigh}}
\end{gather}
where $L_\mathrm{SN}$ is the average energy rate per star during the explosion period.
SPH particles that receive energy turn off radiative cooling calculations and evolve adiabatically.
This technique has first been advocated by \citet{MoriYoshiiTsujimotoNomoto_1997_EvolutionDwarfGalaxies_ApJ}, and then numerous investigations have already scrutinized and sophisticated this technique
\citep[][and so on]{Gerritsen_1997_StarFormationInterstellar_Ph.D.Thesis, 
 MoriYoshiiNomoto_1999_DissipativeProcessMechanism_ApJ, ThackerCouchman_2000_ImplementingFeedbackSimulations_Astrophys.J.}.

\subsubsection{Time stepping}
The simulation time steps $\Delta t$ share the same value throughout the system. It is determined by the CFL conditions: 
\begin{gather}
    \Delta t = \min_i(\Delta t_{i,\text{grav}},\,\Delta t_{i,\text{hydro}}),\\
    \Delta t_{i,\text{grav}}=C_\mathrm{CFL}\sqrt{\frac{\epsilon}{|\mathrm{d}\bm{v}_i/\mathrm{d}t|}},\\
    \Delta t_{i,\text{hydro}} = C_\mathrm{CFL}\frac{h_i}{\max_j(v_{ij}^\mathrm{sig})},
\end{gather}
where $C_\mathrm{CFL}$ is the CFL constant and we set $C_\mathrm{CFL}=0.3$. We adopted the second-order Runge-Kutta method for the time integration.

\subsubsection{Implementation}
This code is parallelized by the Framework for Developing Particle Simulators \citep[\texttt{FDPS}:][]{IwasawaTanikawaHosonoNitadoriEtAl_2016_ImplementationPerformanceFDPS_PublAstronSocJpnNihonTenmonGakkai,NamekataIwasawaNitadoriTanikawaEtAl_2018_FortranInterfaceLayer_PublicationsoftheAstronomicalSocietyofJapan}. In \texttt{FDPS}, the code for parallelization is separated from the code for computing interactions and time integrals. 
It includes functions such as domain decomposition, redistribution of particles, and gathering of particle information for interaction calculation. The \texttt{FDPS} libraries are implemented using OpenMP for intra-node parallelism and MPI for inter-node parallelism. 
Using these libraries, users can implement parallelized programs by writing sequential code for interaction calculations.
The gravitational force is calculated with a tree algorithm \citep{BarnesHut_1986_HierarchicalLogForcecalculation_Nature, Barnes_1990_ModifiedTreeCode_JournalofComputationalPhysics} and the tree-opening angle is $0.7$.
Our code has already been validated by running various test problems, including the shock-tube test, the Evrard collapse and so on \citep{OtakiMori_2022_StudyGalaxyCollisions_Comput.Sci.ItsAppl.-ICCSA2022Workshop}.

\subsection{Initial condition}
In order to study the essential process of galaxy formation for DMSHs collision, we have set up an ideal situation for a head-on collision. Each of the two colliding DMSHs have a total mass of $M_\mathrm{sub}=M_\mathrm{DM}+M_\mathrm{gas}$, containing no stellar components, and the mass ratio between dark matter and gas is $5.36$. 
Each DMSH centres initially at 

\begin{gather}
    (x,\,y,\,z)=(\pm5,\,0,\,0)\,\mathrm{kpc},
\end{gather}
and the initial bulk velocities of the DMSHs are 
\begin{gather}
    (v_x,\,v_y,\,v_z)=(\mp {v_\mathrm{rel}/2},\,0,\,0),
\end{gather}
respectively.
The density distribution of dark matter adopts the NFW profile. The gas is assumed under the hydrostatic equilibrium in the gravitational potential of the dark matter halo,
 \begin{gather}
    \rho_\mathrm{gas}(r)=\rho_\mathrm{gas,0}\exp{\left[-\frac{\mu m_\mathrm{p}}{k_\mathrm{B}T_\mathrm{vir}}\Phi_\mathrm{NFW}(r)\right]},
\end{gather}
where $T_\mathrm{vir}$ is the virial temperature of DMSH defined as
\begin{gather}
    T_\mathrm{vir}=\frac{c(c^2+2c-2(1+c)\ln{(1+c)})}{2((1+c)\ln{(1+c)}-c)^2}\frac{GM_\mathrm{sub}\mu m_\mathrm{p}}{3 k_\mathrm{B}R_{200}}.
\end{gather}
To generate the initial conditions of a DMSH, we use the \texttt{MAGI} \citep{MikiUmemura_2018_MAGIManycomponentGalaxy_MonNotRAstronSoc}. 
After generating particle distributions using \texttt{MAGI}, we calculated a DMSH for several hundred million years in an isolated system of adiabatic processes to suppress density fluctuations and reach dynamical equilibrium.
We run collision simulations of DMSHs with the same mass for three different cases: $M_\mathrm{sub} = 10^8,\,10^9\text{ and } 10^{10}\,\mathrm{M_\odot}$
The number of $N$-body particles and SPH particles is $\sim10^6$, and all particles have the same mass.
We set the gravitational softening length $\epsilon = 0.1\,\mathrm{kpc}$ as the spatial resolution.
The cooling rate for a given metallicity of the gas is calculated by \texttt{MAPPINGS V} \citep{SutherlandDopita_2017_EffectsPreionizationRadiative_ApJS, SutherlandDopitaBinetteGroves_2018_MAPPINGSAstrophysicalPlasma_AstrophysicsSourceCodeLibrary} assuming the Collisional Ionisation Equilibrium (CIE).


\section{results of simulations}
We begin by showing the simulation results for the case of the DMSH collision with a mass of $10^9\,\mathrm{M_\odot}$ and a {relative velocity of $20\,\mathrm{km\,s^{-1}}$}.
Fig. \ref{fig: 1e9Msun_10kms} shows, from top to bottom, the density distribution of dark matter, gas, and star in a thin slice at $z=0$, and the elapsed times are $0,\,285,\,570,\text{ and }884\,\mathrm{Myr}$ from left to right, respectively.


At $285\,\mathrm{Myr}$, the centres of DMSHs collide with each other, compressing the gas and increasing the gas density at the centre. Accordingly, star formation is activated in the central part of the colliding DMSHs.
Shock waves are simultaneously generated at the collision surface and propagate upstreams. 
A high-density gas layer is then formed in the $x=0$ plane, and a large amount of the gas is ejected along this plane.

At $570\,\mathrm{Myr}$, the DMSHs are gravitationally attracted to each other and merge. 
The gravitational contraction of the gas component in the centre of the merged DMSHs induces a burst of star formation. Subsequently, massive stars explode as core-collapse supernovae, heating the surrounding gas to a temperature of $\sim10^6\,\mathrm{K}$. We can observe there is an expanding superbubble driven by the supernova feedback at the left side of the collision surface in the middle panel. As a result, star formation is partially suppressed.
After $500\,\mathrm{Myr}$, the star formation rate is $\sim0.001\,\mathrm{M_\odot \,yr^{-1}}$ and the star formation is stable.
As predicted by the analytical model, two DMSHs merge and form a normal dark-matter-dominated galaxy having dark matter mass $M_\mathrm{DM}=1.16\times10^9\,\mathrm{M_\odot}$, gas mass $2.88\times10^7\,\mathrm{M_\odot}$ and stellar mass $5.19\times10^6\,\mathrm{M_\odot}$ at $4.7\,\mathrm{Gyr}$. 
These are calculated as the masses enclosed within the bound radius, defined as the maximum radius of the stellar particles binding in the system.

Fig. \ref{fig: 1e9Msun_100kms} shows the result of the DMSH collision for a {relative velocity of $200\,\mathrm{km\,s^{-1}}$}. It is the same as Fig. \ref{fig: 1e9Msun_10kms}, but the elapsed times are $0,\,50,\,100,\text{ and }265\,\mathrm{Myr}$ from left to right, respectively.
The centres of each DMSH collide at $50\,\mathrm{Myr}$. On the collision surface of the high-density gas, star formation occurs. 
At $100\,\mathrm{Myr}$, the dark matter components in the DMSHs penetrate each other. 
Then, as the gravitational potential becomes shallower with a short timescale, the distribution of the stellar component expands, decreasing its density.
The stars that form before the collision pass through in a similar way to the motion of dark matter.
Therefore, the stellar component of the galaxies in the collision surface is formed by the collision of the gas in the DMSH. 
{At the same time, several gravitationally bound star clusters with masses of about $10^5\,\mathrm{M_\odot}$ are formed due to the fragmentation of the dense gas layer on the collision surface.}
After $265 \,\mathrm{Myr}$, the dark matter has completely passed through, leaving only a system of gas and stars on the collision surface.
As predicted by the analytical model, dark matter components of DMSHs penetrate through, but gaseous medium collision each other, and enhancement of the gas density induces a burst of star formation on the collision surface. 
Then collided gaseous medium form a dark-matter-deficient galaxy composed with stellar mass $1.34\times10^7\,\mathrm{M_\odot}$ and gas mass $4.02\times10^5\,\mathrm{M_\odot}$ at $3\,\mathrm{Gyr}$.

Fig. \ref{fig: ColDens} shows the stellar surface density of the dark-matter-deficient galaxy. The grey area shows the length below the gravity resolution $\epsilon$.
The black points are stellar surface densities calculated from the star particles projected in the face-on direction.
The curve lines fitted with the S\`{e}rsic profile of the stellar component represented by solid lines. The colour of these lines corresponds to each plane. The effective radii ($r_\mathrm{e}$) and S\`{e}rsic indexes ($n_\text{S\`{e}rsic}$) for the red and green lines are ($r_\mathrm{e},\, n_\text{S\`{e}rsic})=(0.5\,\mathrm{kpc},\,0.8)$.

The top panel in Fig. \ref{fig: SFR_EnclosedMass} shows the evolution of the enclosed masses $M(<r_\mathrm{half})$ within the half mass radius $r_\mathrm{half}=7.1\,\mathrm{kpc}$ from the origin $(x,y,z)=(0,0,0)$ for the {relative velocity of $200\,\mathrm{km\,s^{-1}}$}. $r_\mathrm{half}$ is given by the half mass of a DMSH in the initial condition. The red, blue and green lines are enclosed masses of star, gas and dark matter, respectively.
Since the time of central collision $50\,\mathrm{Myr}$, the enclosed mass of dark matter has been decreasing and the star form from the gas remaining on the collision surface.
Beyond $100\,\mathrm{Myr}$, the gas mass decreases, and at the same time the stellar mass increases.
The star formation history shows the bottom panel in Fig. \ref{fig: SFR_EnclosedMass}. For {a relative velocity of $200\,\mathrm{km\,s^{-1}}$}, the centre of each DMSH collides at $50\,\mathrm{Myr}$, at which time the star formation rate peaks. 
After that, the star formation rate gradually decreases with oscillations due to the alternating enhancement of star formation by radiative cooling and suppression of star formation by heating due to supernova feedback.
After $300\,\mathrm{Myr}$, the gas density decreases due to outflow driven by the supernova feedback, and the star formation rate transitions to a lower state of about several $10^{-4}\,\mathrm{M_\odot\,yr^{-1}}$.

Fig. \ref{fig: RhoT} shows the evolution of the density-temperature diagram for the {relative velocity of $200\,\mathrm{km\,s^{-1}}$}. Colour represents the mass of gas at the specified density and temperature in the simulation box.
The dashed line is the gas density threshold for star formation represented by Equation \eqref{eq: nSF}.
In the initial condition, the gas of the DMSHs is under the virial equilibrium with the virial temperature of $\sim 10^4\,\mathrm{K}$. 

At $50\,\mathrm{Myr}$, the density of the gaseous medium increases due to collisions, and radiative cooling is effective at the centre of the DMSH. 
The low-density gas $(<10^{-26}\,\mathrm{g\,cm^{-3}})$ located on the outskirt of each halo is adiabatically compressed, and its temperature rises to over $10^5\, \mathrm{K}$.

The dense gas $(>10^{-23}\,\mathrm{g\,cm^{-3}})$ has a lower temperature due to radiative cooling and also exceeds the density threshold for star formation conditions.
After $100\,\mathrm{Myr}$, the star formation rate is at its highest, and the supernova feedback heats the dense gas in the star-forming regions.
After $265\,\mathrm{Myr}$, the outflow driven by the strong feedback blows most of the gas away from the system, eventually giving rise to dark-matter-deficient galaxies.

Fig. \ref{fig: 1e9Msun_600kms} shows the result of the DMSH collision for a {relative velocity of $1200\,\mathrm{km\,s^{-1}}$}. 
It is the same as Fig. \ref{fig: 1e9Msun_10kms}, but the elapsed times are $0,\,10,\,50,\text{ and }300\,\mathrm{Myr}$ from left to right, respectively. 

At $10 \,\mathrm{Myr}$ the DMSHs are just after the central collision, the dark matter components are passed through each other, and a few stars form in the dense gas region of the collision surface
at $50 \,\mathrm{Myr}$.
Finally, after $300 \,\mathrm{Myr}$, no galaxy forms because the shock-breakout occurs after the DMSHs collision and most of the gas is ejected from the system without forming stars.


Table \ref{tab: results} lists the initial conditions and results of our collision simulations. 
The effects of supernova feedback on collision-induced galaxy formation are discussed in Section \ref{sec: Feedback}. 
This paper adopts a model in which gas particles evolve adiabatically while receiving thermal energy from supernova explosions as a fiducial feedback model.
This table provides the properties of the most massive galaxy that formed after the DMSH collision. We define $r_\mathrm{bound}$ as the region of the galaxy where the stellar component is gravitationally bound, and the table represents the masses enclosed within that radius.
As a result of the collision simulations, we classify the collision-induced objects as "normal dwarf", "dark-matter-deficient galaxy (DMDG)", "{star} cluster" or "no galaxy". 
We define a galaxy with a dark matter fraction of more than $50$\% ($f_\mathrm{DM}=M_\mathrm{DM}/M_\mathrm{tot}>0.5$) as a dark-matter-dominated galaxy and a galaxy with $f_\mathrm{DM}{\leq}0.5$ as a dark-matter-deficient galaxy. In particular, we defined {star} clusters as objects that have no dark matter at all and a stellar mass of less than $10^6\,\mathrm{M_\odot}$.

The results of the collision simulations are summarised in Fig. \ref{fig: strong[Z]=-1}.
Filled red circles and {blue} squares indicate the formation of a dark-matter-deficient galaxy and a dark-matter-dominated galaxy, respectively. Open red circles are the formation of {star} clusters by the fragmentation at the collision surface. The crosses indicate the result of no galaxy formation at the collision surface.
The solid lines are the results of analytical models; the upper line corresponds to the shock-breakout condition and the lower line to the merger condition. The dashed line is the Jeans mass calculated from the temperature $10^4\,\mathrm{K}$, which is discussed in Section \ref{sec: Feedback}.


\begin{figure}
    \centering
    \includegraphics[width=\columnwidth]{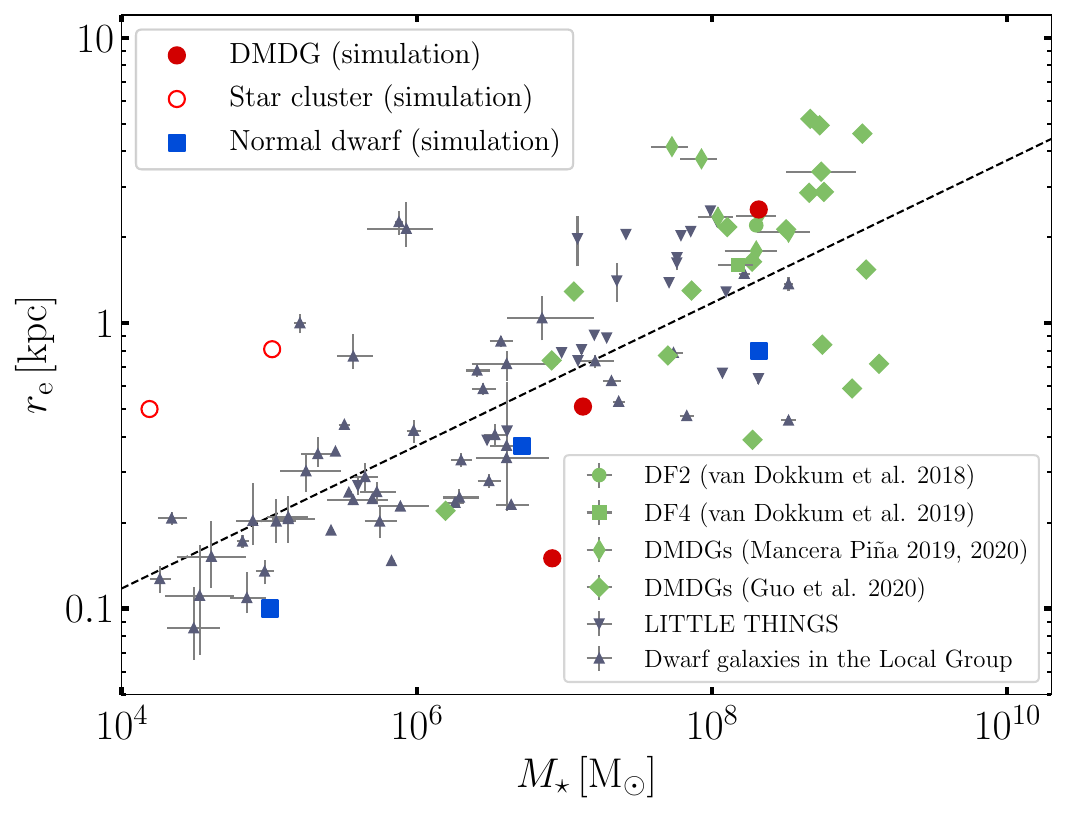}
    \caption{Comparison between observed galaxies and simulated galaxies of the stellar mass versus the effective radius.
    The red {filled circles, red open circles,} and blue squares indicate the dark-matter-deficient galaxies{, star cluster,} and normal dwarf galaxies in our simulations, respectively. The green circle, square, thin diamonds and thick diamonds are observational results of dark-matter-deficient galaxies reported by \citet{vanDokkumDanieliCohenMerrittEtAl_2018_GalaxyLackingDark_Nature}, \citet{DokkumDanieliAbrahamConroyEtAl_2019_SecondGalaxyMissing_ApJL}, \citet{ManceraPinaFraternaliAdamsMarascoEtAl_2019_BaryonicTullyFisher_ApJL, ManceraPinaFraternaliOmanAdamsEtAl_2020_RobustKinematicsGasrich_MNRAS} and \citet{GuoHuZhengLiaoEtAl_2020_FurtherEvidencePopulation_NatAstron}, respectively. The downward and upward triangles represent dwarf galaxies in the catalogue of LITTLE THINGS data \citep{HunterFicut-VicasAshleyBrinksEtAl_2012_LITTLETHINGS_AJ, OhHunterBrinksElmegreenEtAl_2015_HIGHRESOLUTIONMASSMODELS_AJ} and dwarf galaxies in the Local Group \citep{McConnachie_2012_OBSERVEDPROPERTIESDWARF_AJ, BattagliaNipoti_2022_StellarDynamicsDark_NatAstron}, respectively. 
    The dashed line indicates
    the average mass--size relation for dwarf satellites in the local volume given by \citet{CarlstenGreeneGrecoBeatonEtAl_2021_StructuresDwarfSatellites_ApJ}.}
    \label{fig: results_observation}
\end{figure}

\section{Summary and Discussion}

Based on the analytical and numerical studies of the DMSH collisions, galaxy collisions play a significant role in the formation of dark-matter-deficient galaxies. 
{We estimated the distribution of collision frequency in the host galaxy and found that most collisions occur within $0.1\,R_\mathrm{200,\,host}$.
The total collision frequency and collision timescale are $74.2 \,\mathrm{Gyr^{-1}}$ and $13.5\,\mathrm{Myr}$, respectively, for the collisions between DMSHs with $10^9\,\mathrm{M_\odot}$ in the host galaxy with $10^{12}\,\mathrm{M_\odot}$.}
We found the critical {relative} velocities for the bifurcation sequence of the formation of dark-matter-dominated galaxies and dark-matter-deficient galaxies.
The higher {relative} velocities are required to form dark-matter-deficient galaxies in the lower metallicity environments.
In a head-on collision simulation between two DMSHs with the mass of $10^9\,\mathrm{M_\odot}$ including gaseous medium with solar metallicity, a dark-matter-deficient galaxy is formed for the {relative velocity of $200 \,\mathrm{km\,s^{-1}}$}.
A discussion about the detailed physical processes in off-centre collisions of DMSHs will be a future study.
In the following, we compare our results with previous studies and observations. Then, we discuss some physical processes that are not or insufficiently considered in our model.

\subsection{Comparison to the previous studies}
There are several theoretical studies on the formation of dark-matter-deficient galaxies. \cite{Ogiya_2018_TidalStrippingPossible_MNRAS} investigated the formation of dark-matter-deficient galaxies by tidal interaction between a host galaxy and a satellite galaxy using $N$-body simulations. 
In tidal models, tidal tails usually appear on both sides of galaxies, depending on the degree of tidal force. However, observed dark-matter-deficient galaxies do not always have identified clear tidal tails such as NGC1052-DF2. In addition, dark-matter-deficient galaxies are not only found as satellite galaxies of a massive galaxy but also in low-density regions of intergalactic space.
These indicate that tidal models alone do not explain all of the dark-matter-deficient galaxies in terms of galaxy formation scenarios. 
Since there are no restrictions on galaxy morphology in the collision model, as in the tidal model, there will be more situations in which the collision model can be applied. On the other hand, for dark-matter-deficient galaxies in low-density environments, further studies of collision probabilities in such environments are needed.

\cite{ShinJungKwonKimEtAl_2020_DarkMatterDeficient_ApJ} worked on an important study about the formation model of dark-matter-deficient galaxies. In \cite{ShinJungKwonKimEtAl_2020_DarkMatterDeficient_ApJ}, simulations have been carried out in off-centre collisions, which we do not take into account in this paper. Our study has clarified the fundamental physical processes of the DMSH collisions and derived the critical relative velocities for the bifurcation sequence of the formation of dark-matter-dominated galaxies and dark-matter-deficient galaxies. They mentioned excessive supersonic turbulence as a reason dark-matter-deficient galaxies do not form in very high-velocity collisions. On the other hand, our simulations show that gas ejection induced by shock breakouts is essential to suppress the formation of dark matter-depleted galaxies. A quantitative and detailed comparison of the differences between these claims will be necessary.

\cite{MadauLupiDiemandBurkertEtAl_2020_GlobularClusterFormation_ApJ} study a scenario for the formation of globular clusters (GCs) triggered by fast collisions between DMSHs. It is interesting to note that the extrapolation of our analytical model (Fig.\ref{fig: 1Dresult}) to the low-mass side may provide insight into this GC formation model. Furthermore, our simulations also show that the growth of instabilities generated at the collision surface can lead to the fragmentation of high-density regions, resulting mass of star clusters as high as those of observed globular clusters. It would be fascinating to investigate whether these meet the observational properties of globular clusters; however, this is still difficult due to the numerical resolution in our current simulations. Therefore, future high-resolution calculations are expected.


We show the galaxies formed in our collision simulations and the observed normal dwarf and dark-matter-deficient galaxies in Fig. \ref{fig: results_observation}. 
The red circles and blue squares indicate the dark-matter-deficient galaxies and normal dwarf galaxies in our simulations, respectively. The green circle, square, thin diamonds and thick diamonds are observational results of dark-matter-deficient galaxies reported by \citet{vanDokkumDanieliCohenMerrittEtAl_2018_GalaxyLackingDark_Nature}, \citet{DokkumDanieliAbrahamConroyEtAl_2019_SecondGalaxyMissing_ApJL}, \citet{ManceraPinaFraternaliAdamsMarascoEtAl_2019_BaryonicTullyFisher_ApJL, ManceraPinaFraternaliOmanAdamsEtAl_2020_RobustKinematicsGasrich_MNRAS} and \citet{GuoHuZhengLiaoEtAl_2020_FurtherEvidencePopulation_NatAstron}, respectively. The downward and upward triangles represent dwarf galaxies in the catalogue of LITTLE THINGS data \citep{HunterFicut-VicasAshleyBrinksEtAl_2012_LITTLETHINGS_AJ, OhHunterBrinksElmegreenEtAl_2015_HIGHRESOLUTIONMASSMODELS_AJ} and dwarf galaxies in the Local Group \citep{McConnachie_2012_OBSERVEDPROPERTIESDWARF_AJ, BattagliaNipoti_2022_StellarDynamicsDark_NatAstron}, respectively.
Fig. \ref{fig: results_observation} indicates the stellar mass--size ($M_\star\text{--}r_\mathrm{e}$) relation for the galaxies. The dashed line indicates the average mass--size relation for dwarf satellites in the local volume given by \citet{CarlstenGreeneGrecoBeatonEtAl_2021_StructuresDwarfSatellites_ApJ}. UDGs are defined as galaxies with effective radii greater than $1.5\,\mathrm{kpc}$. The collision simulation between $10^{10}\,\mathrm{M_\odot}$ DMSHs shows the formation of a single dark-matter-deficient galaxy with $r_\mathrm{e}=2.5\,\mathrm{kpc}$ at a {relative velocity of $400\,\mathrm{km\,s^{-1}}$}. {Therefore, this dark-matte-deficient galaxy formed by such the collision process could be observed as a UDG.}
On the other hand, dwarf galaxies formed at slower {relative} velocities show a slight offset to the minor side in the mass-radius relationship derived from observations. This might be due to the fact that the supernova feedback, which changes the gravitational potential through galactic outflows, still has only a little effect.

\subsection{Effects of supernova feedback}\label{sec: Feedback}

It is well known that subgrid models of supernova feedback have a significant impact on galaxy formation simulations. 
Supernova feedback heats up the ambient gas and decreases the star formation rate of the galaxy through gas outflows.
Recently, various methods have been developed to give supernova feedback in a more appropriate way \cite[e.g., ][]{DallaVecchiaSchaye_2012_SimulatingGalacticOutflows_MonthlyNoticesoftheRoyalAstronomicalSociety,ShimizuTodorokiYajimaNagamine_2019_OsakaFeedbackModel_MonNotRAstronSoc, OkuTomidaNagamineShimizuEtAl_2022_OsakaFeedbackModel_ApJS}. 
This strong feedback causes a large amount of the interstellar medium to be ejected out of the system, subsequently causing the gravitational potential of the system to become shallower. If this occurs on a timescale sufficiently shorter than the dynamical time of the system, the system expands quickly and achieves a new dynamical equilibrium state.
Much work has been done on these fundamental physical processes, with analytical treatments investigated by \citep{DekelSilk_1986_OriginDwarfGalaxies_TheAstrophysicalJournal} and demonstrated by \citet{MoriYoshiiTsujimotoNomoto_1997_EvolutionDwarfGalaxies_ApJ, MoriYoshiiNomoto_1999_DissipativeProcessMechanism_ApJ} in numerical simulations. More recently, 
\citep{DiCintioBrookDuttonMaccioEtAl_2017_NIHAOXIFormation_MonthlyNoticesoftheRoyalAstronomicalSociety:Letters} analysed its effects on UDGs

In the case of effective cooling, the temperature of primordial gas could be $10^4\,\mathrm{K}$ in CIE. We consider the collisions of isothermal gas clouds and isothermal shock in our analytical model. The conditions are the same as the adiabatic shock-breakout condition, but the isothermal shock of $10^4\,\mathrm{K}$ is generated at the collision surface. 
The gas density of shocked clouds are 
\begin{gather}
    \rho_{1,\,\mathrm{iso}}= \frac{\mathcal{M}^2+\mathcal{M}\sqrt{\mathcal{M}^2+4}+2}{2}\rho_0,
\end{gather}
where $\mathcal{M}=v_\mathrm{rel}/(2c_{\mathrm{s,\,iso}})$ is the isothermal Mach number and  $c_{\mathrm{s,\,iso}}=8.2\,\mathrm{km\,s^{-1}}$ is the isothermal sound speed for gas metallicity $0.1\,\mathrm{Z_\odot}$. The Jeans instability criterion in two gas clouds is 
\begin{gather}
    2\beta M_\mathrm{gas}\geq M_\mathrm{J}\equiv \sqrt{\frac{\pi^5 c_\mathrm{s,\,iso}^6}{36G^3\rho_{1,\,\mathrm{iso}}}},
\end{gather}
where $\beta$ is the parameter. In Fig. \ref{fig: strong[Z]=-1}, the dashed lines indicate the Jeans instability criteria for $\beta = 1.0,\, 0.1,\, 0.01$, from left to right, respectively. In collision simulations between subhaloes with masses of $10^8\,\mathrm{M_\odot}$, no galaxies formed on the collision surface at { relative velocities of $20,\,100,\text{ and }200\,\mathrm{km\,s^{-1}}$} since gaseous medium in subhaloes are not Jeans unstable.
On the other hand, for {relative velocities of $400,\,600,\text{ and }800\,\mathrm{km\,s^{-1}}$}, the gas fragments and star formation occur after DMSH collisions inducing the formation of star clusters with less than $1/10$ of the initial gas mass.

\begin{figure}
    \centering
    \includegraphics[width=\columnwidth]{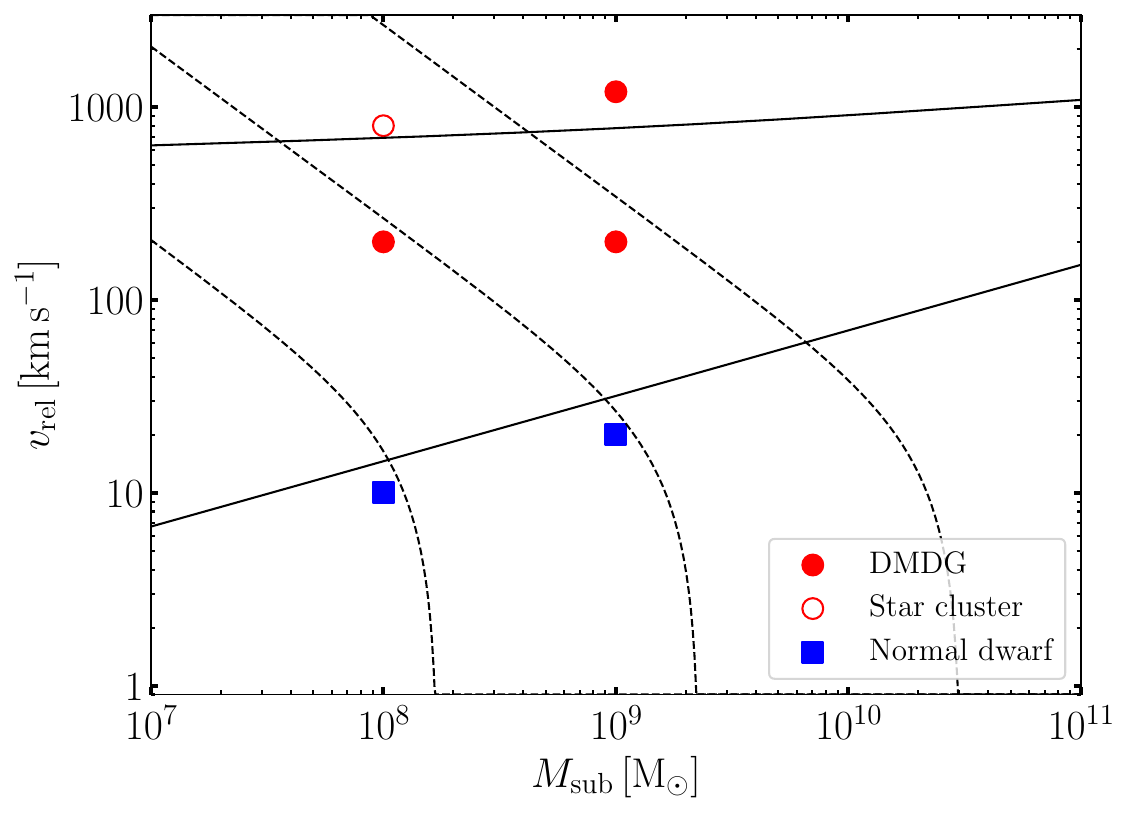}
    \caption{Same as Figure \ref{fig: strong[Z]=-1}, but results for the weak feedback model.}
    \label{fig: weak[Z]=-1}
\end{figure}

Next, we run collision simulations for the other feedback model to compare the formation processes and the property of collision-induced galaxy. This model is implemented as the SPH particles receive feedback energy from supernova and evolve without turning off radiative cooling calculations. 
Since the effect of supernova feedback is weak for gas heating, the gas outflow rate is lower than in previous simulations. 
{Based on this weak feedback model, we simulate collisions between DMSHs with the same masses ($10^8\,\mathrm{M_\odot}$ or $10^9\,\mathrm{M_\odot}$) for three relative velocities (low, moderate, and high speed), respectively. All other parameters and initial conditions are the same as the strong feedback model.}
We summarise the result of simulations for this weak feedback model in Table \ref{tab: results} and Fig. \ref{fig: weak[Z]=-1}. 
{Compared to the strong feedback model, simulated galaxies in the weak feedback model evolve with a higher star formation rate on average, although the maximum star formation rate at the time of collision is comparable. Moreover, the low gas outflow rate due to the weak feedback results in the formation of galaxies and star clusters with massive stellar masses and smaller effective radii. In the case of a collision simulation of $10^9\,\mathrm{M_\odot}$ with a relative velocity of $200\,\mathrm{km\,s^{-1}}$, a DMDG is formed with a stellar mass of $1.34\times10^7\,\mathrm{M_\odot}$ and $1.99\times10^8\,\mathrm{M_\odot}$ and an effective radius of $0.51\,\mathrm{kpc}$ and $0.092\,\mathrm{kpc}$ in the strong and weak feedback models, respectively. 
Since the feedback model has a significant effect on the properties of galaxies formed by collisions, it is very interesting that future observations of these galaxies evaluate how effectively the supernova feedback influences the formation of these galaxies.}

\subsection{Radiative cooling}

\begin{figure*}
  \begin{minipage}[b]{\columnwidth}
    \centering
    \includegraphics[width=\columnwidth]{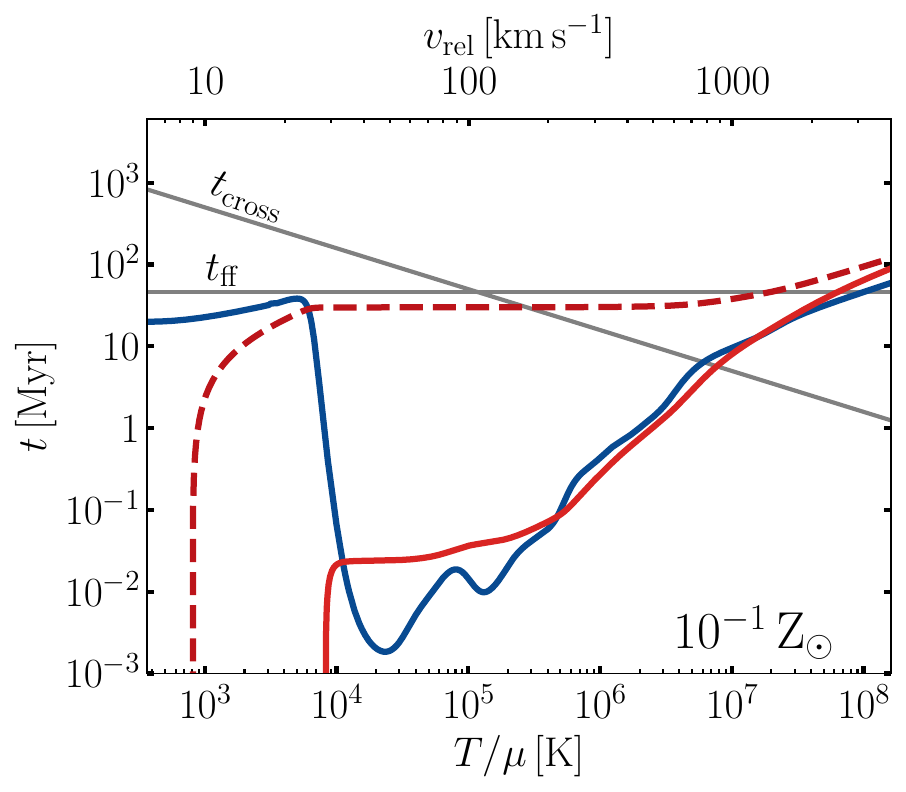}
  \end{minipage}
  \hspace{0.04\columnwidth}
  \begin{minipage}[b]{\columnwidth}
    \centering
    \includegraphics[width=\columnwidth]{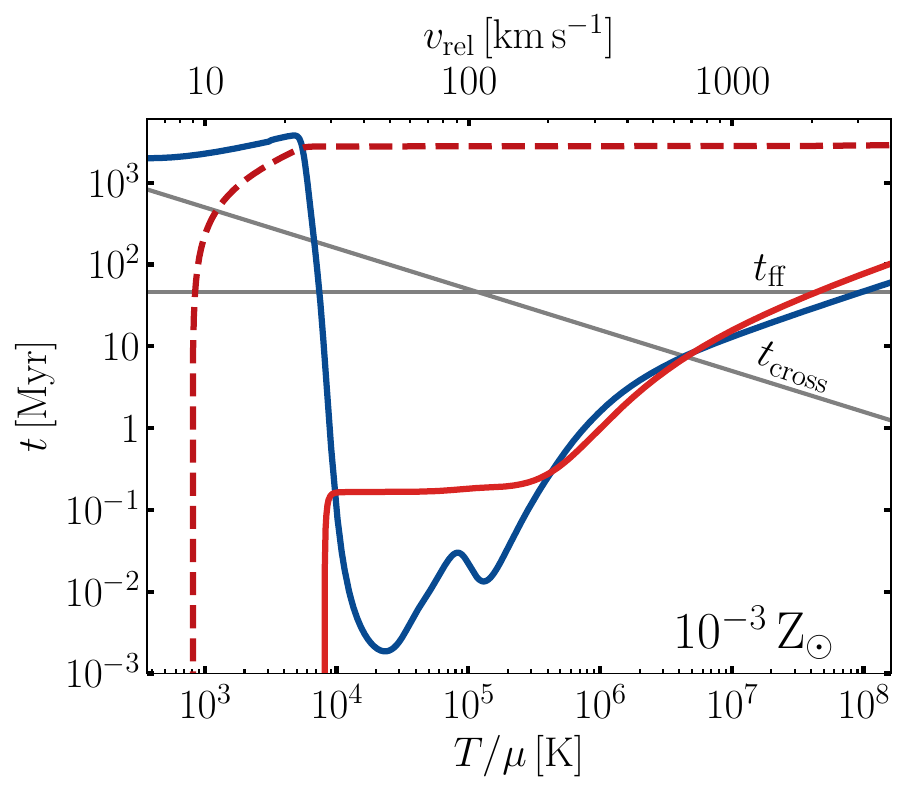}
  \end{minipage}
    \caption{Timescales after collision between DMSHs with $10^9\,\mathrm{M_\odot}$ in the analytical model. The left panel and right panel are the gas metallicity for $10^{-1}\,\mathrm{Z_\odot}$ and $10^{-3}\,\mathrm{Z_\odot}$, respectively. The temperature {divided by the mean molecular weight $T/\mu$} corresponds to the kinetic energy of the {relative velocity $v_\mathrm{rel}$}. The blue line is the conventional cooling time $t_\mathrm{conv,\,eff}$, the red lines are the effective cooling time $t_\mathrm{cool,\,eff}$ with the solid line corresponding to the timescale for $T_\mathrm{end}=10^4\,\mathrm{K}$, and the dashed line corresponding to the timescale for $T_\mathrm{end}=10^3\,\mathrm{K}$. The two grey lines are the shock-crossing time $t_\mathrm{cross}$ and the free-fall time $t_\mathrm{ff}$, respectively.}
    \label{fig: EffectiveCoolingTime}
\end{figure*}

Throughout this paper, we use the EI scheme \citep{Townsend_2009_EXACTINTEGRATIONSCHEME_ApJS} to calculate the radiative cooling term of the energy equation in the simulations. 
In the EI scheme, the time evolution of the radiative cooling term in the energy equation can be solved for its temperature dependence by fitting a temporal evolution function $Y(T)$ integrating the inverse of the cooling rate. 
In order to consider the thermodynamic evolution of the gas in DMSH collision simulations, we define the cooling time using the EI scheme, instead of the conventional cooling time $t_\mathrm{cool,\,conv}$ given by the equation \eqref{eq: CoolingTime}. In the following, the temperature is used instead of the specific internal energy.
The effective cooling time is defined as 
\begin{align}
    t_\mathrm{cool,\,eff}(u_\mathrm{start}\rightarrow u_\mathrm{end})
    &= -\frac{m_\mathrm{p}^2}{\rho}\int_{u_\mathrm{start}}^{u_\mathrm{end}}\frac{\mu(u)^2}{\Lambda(u,Z)}\mathrm{d}u,\\
    &
    =Y(u_\mathrm{end}) \, t_\mathrm{cool, \,conv} (u_\mathrm{start}).
\end{align} 
This means the cooling time given for the energy $u_\mathrm{start}$ to cool down to $u_\mathrm{end}$. 

Fig. \ref{fig: EffectiveCoolingTime} illustrates the timescales after a collision between DMSHs with $10^9 \,\mathrm{M_\odot}$ in the analytical model (\ref{sec: ShockBreakout}) as a function of the relative velocity. The left and right panels correspond for the gas metallicity for $Z=10^{-1} \,Z_\odot$ and $Z=10^{-3} \,Z_\odot$, respectively. The panels display the conventional cooling time $t_\mathrm{cool, conv}$ as solid red lines, while the effective cooling time $t_\mathrm{cool, eff}$ is represented by dashed red lines. The free--fall time 
\begin{gather}
t_\mathrm{ff} = \sqrt{\frac{3\pi}{32 G \rho}},
\end{gather}
is depicted as almost horizontal solid lines, and the remaining black solid lines denote the shock-crossing time $t_\mathrm{cross}$. These timescales are represented as a function of the gas temperature after the collision, assuming that the kinetic energy is entirely converted to internal energy. 
In the case of DMSHs colliding with a velocity of $100 \,\mathrm{km \,s}^{-1}$ for $Z=10^{-1} \,Z_\odot$, the conventional cooling time at a temperature of $2.4\times10^5 \,\mathrm{K}$ is about $0.02 \,\mathrm{Myr}$. In contrast, the effective cooling time for $T_\mathrm{end} = 10^3 \,\mathrm{K}$ is $10 \,\mathrm{Myr}$. The conventional definition yields a shorter cooling time than the new definition because of the temperature dependence of the cooling rate. Consequently, the present method is significantly different from the conventional method, and it is highly effective in accurately tracking radiative cooling in the study of galaxy formation.



Another important physical process in galaxy evolution is molecular cooling.
This process is effective at low metallicity of $10^{-3}\,\mathrm{Z}_\odot$. To study the evolution of galaxies without the gravitational potential of dark matter, it is necessary to solve molecular cooling and non-equilibrium chemical calculations. 
However, in this paper, molecular cooling does not come into play since we run collision simulations between DMSHs with metal abundances of $10^{-1}\,\mathrm{Z}_\odot$
For low-temperature gases below $10^4 \,\mathrm{K}$, ignoring the effects of dust, it is known that cooling by heavy elements is more efficient than molecular hydrogen cooling for gases containing this amount of heavy elements.

\subsection{Thermal conduction}

\begin{figure}
    \centering
    \includegraphics[width=\columnwidth]{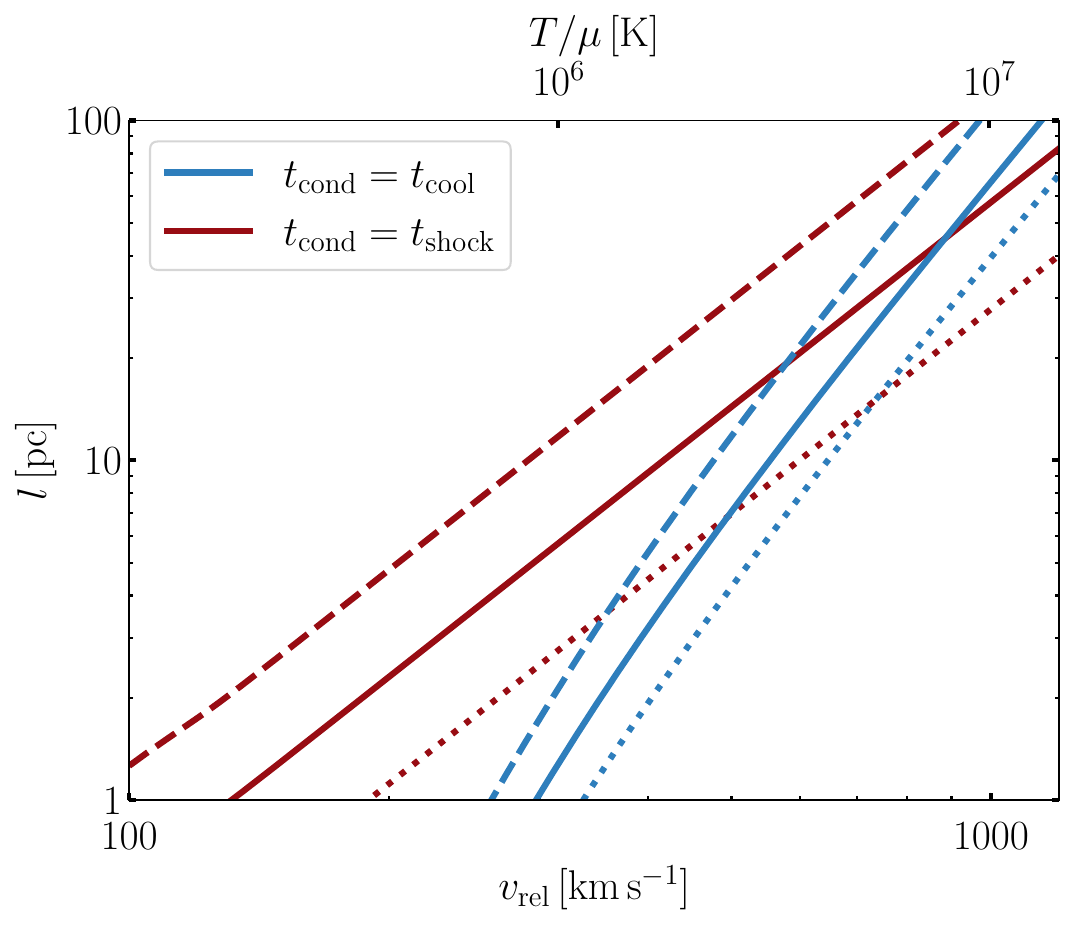}
    \caption{Scale length of the temperature gradient for the {relative} velocity of DMSH in our analytical model. The blue and red lines indicate that the thermal conduction timescales equal the cooling time and shock crossing time, respectively. medium with $0.1\,\mathrm{Z_\odot}$. 
    The dotted, solid and dashed lines correspond to the subhalo mass of $10^8\,\mathrm{M_\odot},\, 10^9\,\mathrm{M_\odot}$ and $10^{10}\,\mathrm{M_\odot}$ assuming $M_\mathrm{DM}/M_\mathrm{gas}=5.36$, respectively.}
    \label{fig: ConductionTimeScale}
\end{figure}

We use the analytical model to estimate the effect of thermal conduction on the physical processes of DMSH collisions.
The spatial resolution is $100 \,\mathrm{pc}$ for all simulation results in this paper, but a higher resolution is needed to compare the result of simulations with the observed compact dwarf galaxies 
and globular clusters. 
Since the thermal conduction by electrons is also an important physical process in plasma physics on a small scale, we use the analytical model to estimate the effect of thermal conduction on the physical processes of galaxy collisions.

The time scale of thermal conduction is defined as 
\begin{gather}
    t_\mathrm{cond}=\frac{\rho k_\mathrm{B}l^2}{(\gamma-1)\mu m_\mathrm{p}\kappa},
\end{gather}
where $l$ is the scale length of the temperature gradient. The thermal conductivity for a hydrogen plasma $\kappa$ is given by \citet{CowieMcKee_1977_EvaporationSphericalClouds_Astrophys.J.},
\begin{gather}
    \kappa(T)=1.31\,\frac{n_\mathrm{e}\lambda k_\mathrm{B}^{3/2}T^{1/2}}{m_\mathrm{e}^{1/2}},
\end{gather}
where $m_\mathrm{e}$ is the electron mass, $n_e$ is the electron density and $\lambda$ is the equivalent mean free path for electrons,
\begin{gather}
    \lambda=\frac{3^{3/2}(k_\mathrm{B}T_\mathrm{e})^{2}}{4\pi^{1/2}n_\mathrm{e}e^4\ln\Lambda},
\end{gather}
where $T_\mathrm{e}$ is the electron temperature, $e$ is the elementary charge, and the Coulomb logarithm $\ln\Lambda $ is 
\begin{gather}
    \ln\Lambda = 37.8+\ln\left[\left(\frac{T_\mathrm{e}}{10^8\,\mathrm{K}}\right)\left(\frac{n_\mathrm{e}}{10^{-3}\,\mathrm{cm^{-3}}}\right)^{-1}\right].
\end{gather}
We assume the $T_\mathrm{e}=T$ since the equilibrium timescale of electrons and ions is shorter than the cooling time in this model.

Fig. \ref{fig: ConductionTimeScale} represents the scale length of the temperature gradient $l$ as functions of the {relative velocity $v_\mathrm{rel}$}, corresponding to the case in which the cooling time (equation \eqref{eq: CoolingTime}) or shock crossing time (equation \eqref{eq: ShockTime}) is equal to the thermal conduction time, respectively. 
The gas temperature in the post-collision is calculated on the assumption that the kinetic energy of {relative} velocities between subhaloes is converted to the internal energy of gaseous medium with $0.1\,\mathrm{Z_\odot}$. 
The dotted, solid and dashed lines correspond to the subhalo mass of $10^8\,\mathrm{M_\odot},\, 10^9\,\mathrm{M_\odot}$ and $10^{10}\,\mathrm{M_\odot}$ assuming $M_\mathrm{DM}/M_\mathrm{gas}=5.36$, respectively.
It is clear that since the scale length of the temperature gradient is smaller than the spatial resolution of the simulation $100 \,\mathrm{pc}$, the conduction time is smaller than the cooling time and the shock crossing time. 
However, the scale length of the temperature gradient can be longer than the numerical resolution when performing high-resolution collision simulations, such as resolving to the several $\mathrm{pc}$, the effective radius of a globular cluster.

The thermal conduction timescale is shorter than the shock crossing time at the velocity {$\sim200\,\mathrm{km\,s^{-1}}$} for the formation of dark-matter-deficient galaxies. 
Thermal conduction may affect the gas outflow rate and the evolution of galaxies.
Therefore, the physical process of thermal conduction needs to be taken into account in high-resolution simulations of DMSH collisions, since it may affect the gas outflow rate and the galaxy evolution and formation.



\begin{landscape}
\begin{table}
\centering
\caption{Results of simulations: total mass of subhalo $M_\mathrm{sub}$, {relative velocity $v_\mathrm{rel}$} for the initial condition of collision simulations, feedback model (Strong or Week) as simulation set-up, enclosed stellar mass $M_\star$, gas mass $M_\mathrm{gas}$, dark matter mass $M_\mathrm{DM}$ within bound radius $r_\mathrm{bound}$, effective radius of a galaxy $r_\mathrm{e}$, S\'ersic index $n_\text{S\'ersic}$ for a most massive galaxy formed via a collision process{, and types of collision-induced galaxies (normal dwarf galaxies with $M_\mathrm{DM}/M_\mathrm{tot}>0.5$, DMDGs with $M_\mathrm{DM}/M_\mathrm{tot}\leq0.5$ and $M_\star \geq 10^6\,\mathrm{M_\odot}$ and star clusters with  $M_\mathrm{DM}/M_\mathrm{tot}\leq0.5$ and $M_\star < 10^6\,\mathrm{M_\odot}$)}.}
\label{tab: results}
\begin{threeparttable}
\begin{tabular}{lccccccccc}\toprule
\multicolumn{2}{c}{Initial conditions} &
  \multicolumn{1}{l}{} &
  \multicolumn{6}{c}{Properties of a most massive galaxy or subhalo\tnote{1}} &
  \multirow{2}{*}{\begin{tabular}[c]{@{}c@{}}Collision-induced objects\end{tabular}} \\
$M_\mathrm{sub}\,\mathrm{[M_\odot]}$ &
  $v_\mathrm{rel}\,\mathrm{[km\,s^{-1}]}$ &
  Feedback &
  $M_\star\,\mathrm{[M_\odot]}$ &
  $M_\mathrm{gas}\,\mathrm{[M_\odot]}$ &
  $M_\mathrm{DM}\,\mathrm{[M_\odot]}$ &
  $r_\mathrm{bound}\,\mathrm{[kpc]}$ &
  $r_\mathrm{e}\,\mathrm{[kpc]}$ &
  $n_\text{S\'ersic}$ &
   \\ \toprule
\multirow{10}{*}{$10^8$}   & {$10$}   & S & $1.02\times10^5$ & $5.88\times10^3$ & $1.77\times10^6$ & $0.2$  & $0.10$  & $0.23$ & Normal dwarf \\
                           & {$10$}   & W & $2.47\times10^7$ & $6.73\times10^5$ & $3.89\times10^7$ & $1.3$  & $0.072$ & $1.8$  & Normal dwarf \\
                           & {$20$}   & S & -                & -                & -                & -      & -       & -      & No galaxy           \\
                           & {$100$}  & S & -                & -                & -                & -      & -       & -      & No galaxy           \\
                           & {$200$}  & S  & -                & -                & -                & -      & -       & -      & No galaxy          \\
                           & {$200$}  & W & $1.56\times10^7$ & $7.94\times10^5$ & $0$              & $15.4$ & $0.011$ & $4.6$  & DMDG\\
                           & {$400$}  & S & $3.68\times10^3$ & $0$              & $0$              & $3.0$  & -       & -      & {Star cluster}\\
                           & {$600$}  & S & $1.05\times10^5$ & $0$              & $0$              & $3.8$  & $0.81$  & $0.3$  & {Star cluster}\\
                           & {$800$}  & S & $1.55\times10^4$ & $0$              & $0$              & $3.3$  & $0.50$  & $0.99$ & {Star cluster}   \\
                           & {$800$}  & W & $7.23\times10^5$ & $6.35\times10^5$ & $0$              & $9.4$  & $0.071$ & $3.11$ & {Star cluster}   \\ \midrule[0.02em]
\multirow{6}{*}{$10^9$}    & {$20$}   & S & $5.19\times10^6$ & $2.88\times10^7$ & $1.16\times10^9$ & $16.1$ & $0.37$  & $6.0$  & Normal dwarf \\
                           & {$20$}   & W & $2.23\times10^8$ & $5.17\times10^6$ & $4.07\times10^8$ & $2.4$  & $0.16$  & $2.2$  & Normal dwarf \\
                           & {$200$}  & S & $1.34\times10^7$ & $4.02\times10^5$ & $0$              & $17.7$ & $0.51$  & $0.83$ & DMDG \\
                           & {$200$}  & W & $1.99\times10^8$ & $9.16\times10^6$ & $0$              & $22.6$ & $0.092$ & $1.4$  & DMDG         \\
                           & {$1200$} & S & -                & -                & -                & -      & -       & -      & No galaxy           \\
                           & {$1200$} & W & $4.83\times10^6$ & $1.48\times10^6$ & $0$              & $21.6$ & -       & -      & DMDG         \\ \midrule[0.02em]
\multirow{3}{*}{$10^{10}$} & {$100$}  & S & $2.08\times10^8$ & $3.58\times10^6$ & $3.29\times10^9$ & $51.2$ & $0.80$  & $11$   & Normal dwarf \\
                           & {$400$}  & S & $2.08\times10^8$ & $6.09\times10^6$ & $0$              & $51.1$ & $2.5$   & $8.9$  & DMDG       \\
                           & {$1200$} & S & $8.30\times10^6$ & $0$              & $0$              & $22.6$ & $0.15$  & $4.1$  & DMDG  \\ \bottomrule
\end{tabular}
\begin{tablenotes}
\item[1] {The dashes mean that no galaxies form or that the fitting error of S\'ersic profile is too large.}
\end{tablenotes}
\end{threeparttable}
\end{table}
\end{landscape}

\section*{Acknowledgments}
{We would like to thank the anonymous referee for the useful suggestions.}
Numerical computations were performed with computational resources provided by the Multidisciplinary Cooperative Research Program in the Center for Computational Sciences, the University of Tsukuba, Oakforest-PACS operated by the Joint Center for Advanced High-Performance Computing (JCAHPC), and the FUJITSU Supercomputer PRIMEHPC FX1000 and FUJITSU Server PRIMERGY GX2570 (Wisteria/BDEC-01) at the Information Technology Center, the University of Tokyo.
This work was supported by
JSPS KAKENHI Grant Numbers 
JP22KJ0370,
JP21J21888, JP20K04022.
\section*{Data Availability}
Data related to this work will be shared on reasonable request to the corresponding author.




\bibliographystyle{mnras}
\bibliography{library}

\begin{thebibliography}{}
\makeatletter
\relax
\def\mn@urlcharsother{\let\do\@makeother \do\$\do\&\do\#\do\^\do\_\do\%\do\~}
\def\mn@doi{\begingroup\mn@urlcharsother \@ifnextchar [ {\mn@doi@}
  {\mn@doi@[]}}
\def\mn@doi@[#1]#2{\def\@tempa{#1}\ifx\@tempa\@empty \href
  {http://dx.doi.org/#2} {doi:#2}\else \href {http://dx.doi.org/#2} {#1}\fi
  \endgroup}
\def\mn@eprint#1#2{\mn@eprint@#1:#2::\@nil}
\def\mn@eprint@arXiv#1{\href {http://arxiv.org/abs/#1} {{\tt arXiv:#1}}}
\def\mn@eprint@dblp#1{\href {http://dblp.uni-trier.de/rec/bibtex/#1.xml}
  {dblp:#1}}
\def\mn@eprint@#1:#2:#3:#4\@nil{\def\@tempa {#1}\def\@tempb {#2}\def\@tempc
  {#3}\ifx \@tempc \@empty \let \@tempc \@tempb \let \@tempb \@tempa \fi \ifx
  \@tempb \@empty \def\@tempb {arXiv}\fi \@ifundefined
  {mn@eprint@\@tempb}{\@tempb:\@tempc}{\expandafter \expandafter \csname
  mn@eprint@\@tempb\endcsname \expandafter{\@tempc}}}

\bibitem[\protect\citeauthoryear{Aghanim et~al.,}{Aghanim
  et~al.}{2020}]{AghanimAkramiAshdownAumontEtAl_2020_Planck2018Results_A&Aa}
Aghanim N.,  et~al., 2020, \mn@doi [A\&A] {10.1051/0004-6361/201833910}, 641,
  A6

\bibitem[\protect\citeauthoryear{Balsara}{Balsara}{1995}]{Balsara_1995_NeumannStabilityAnalysis_JournalofComputationalPhysics}
Balsara D.~S.,  1995, \mn@doi [Journal of Computational Physics]
  {10.1016/S0021-9991(95)90221-X}, 121, 357

\bibitem[\protect\citeauthoryear{Barnes}{Barnes}{1990}]{Barnes_1990_ModifiedTreeCode_JournalofComputationalPhysics}
Barnes J.~E.,  1990, \mn@doi [Journal of Computational Physics]
  {10.1016/0021-9991(90)90232-P}, 87, 161

\bibitem[\protect\citeauthoryear{Barnes \& Hut}{Barnes \&
  Hut}{1986}]{BarnesHut_1986_HierarchicalLogForcecalculation_Nature}
Barnes J.,  Hut P.,  1986, \mn@doi [Nature] {10.1038/324446a0}, 324, 446

\bibitem[\protect\citeauthoryear{Battaglia \& Nipoti}{Battaglia \&
  Nipoti}{2022}]{BattagliaNipoti_2022_StellarDynamicsDark_NatAstron}
Battaglia G.,  Nipoti C.,  2022, \mn@doi [Nat Astron]
  {10.1038/s41550-022-01638-7}, 6, 659

\bibitem[\protect\citeauthoryear{Behroozi, Wechsler  \& Conroy}{Behroozi
  et~al.}{2013}]{BehrooziWechslerConroy_2013_AVERAGESTARFORMATION_ApJ}
Behroozi P.~S.,  Wechsler R.~H.,   Conroy C.,  2013, \mn@doi [ApJ]
  {10.1088/0004-637X/770/1/57}, 770, 57

\bibitem[\protect\citeauthoryear{Binney \& Tremaine}{Binney \&
  Tremaine}{2008}]{BinneyTremaine_2008_GalacticDynamicsSecond_GalacticDynamics:SecondEditionbyJamesBinneyandScottTremaine.ISBN978-0-691-13026-2HB.PublishedbyPrincetonUniversityPressPrincetonNJUSA2008.}
Binney J.,  Tremaine S.,  2008, Galactic Dynamics: Second Edition, by James
  Binney and Scott Tremaine. ISBN 978-0-691-13026-2 (HB). Published by
  Princeton University Press, Princeton, NJ USA, 2008.

\bibitem[\protect\citeauthoryear{Bullock, Kolatt, Sigad, Somerville, Kravtsov,
  Klypin, Primack  \& Dekel}{Bullock
  et~al.}{2001}]{BullockKolattSigadSomervilleEtAl_2001_ProfilesDarkHaloes_MonthlyNoticesoftheRoyalAstronomicalSociety}
Bullock J.~S.,  Kolatt T.~S.,  Sigad Y.,  Somerville R.~S.,  Kravtsov A.~V.,
  Klypin A.~A.,  Primack J.~R.,   Dekel A.,  2001, \mn@doi [Monthly Notices of
  the Royal Astronomical Society] {10.1046/j.1365-8711.2001.04068.x}, 321, 559

\bibitem[\protect\citeauthoryear{Buzzo, Forbes, Brodie, Janssens, Couch,
  Romanowsky  \& Gannon}{Buzzo
  et~al.}{2023}]{BuzzoForbesBrodieJanssensEtAl_2023_LargescaleStructureGlobular_Mon.Not.R.Astron.Soc.}
Buzzo M.~L.,  Forbes D.~A.,  Brodie J.~P.,  Janssens S.~R.,  Couch W.~J.,
  Romanowsky A.~J.,   Gannon J.~S.,  2023, \mn@doi [Monthly Notices of the
  Royal Astronomical Society] {10.1093/mnras/stad1012}, 522, 595

\bibitem[\protect\citeauthoryear{Carlsten, Greene, Greco, Beaton  \&
  {Kado-Fong}}{Carlsten
  et~al.}{2021}]{CarlstenGreeneGrecoBeatonEtAl_2021_StructuresDwarfSatellites_ApJ}
Carlsten S.~G.,  Greene J.~E.,  Greco J.~P.,  Beaton R.~L.,   {Kado-Fong} E.,
  2021, \mn@doi [ApJ] {10.3847/1538-4357/ac2581}, 922, 267

\bibitem[\protect\citeauthoryear{Chhatkuli, Paudel, Bachchan, Aryal  \&
  Yoo}{Chhatkuli
  et~al.}{2023}]{ChhatkuliPaudelBachchanAryalEtAl_2023_FormingBlueCompact_MonthlyNoticesoftheRoyalAstronomicalSociety}
Chhatkuli D.~N.,  Paudel S.,  Bachchan R.~K.,  Aryal B.,   Yoo J.,  2023,
  \mn@doi [Monthly Notices of the Royal Astronomical Society]
  {10.1093/mnras/stac3700}, 520, 4953

\bibitem[\protect\citeauthoryear{Cowie \& McKee}{Cowie \&
  McKee}{1977}]{CowieMcKee_1977_EvaporationSphericalClouds_Astrophys.J.}
Cowie L.~L.,  McKee C.~F.,  1977, \mn@doi [The Astrophysical Journal]
  {10.1086/154911}, 211, 135

\bibitem[\protect\citeauthoryear{Dalla~Vecchia \& Schaye}{Dalla~Vecchia \&
  Schaye}{2012}]{DallaVecchiaSchaye_2012_SimulatingGalacticOutflows_MonthlyNoticesoftheRoyalAstronomicalSociety}
Dalla~Vecchia C.,  Schaye J.,  2012, \mn@doi [Monthly Notices of the Royal
  Astronomical Society] {10.1111/j.1365-2966.2012.21704.x}, 426, 140

\bibitem[\protect\citeauthoryear{Danieli, van Dokkum, Abraham, Conroy, Dolphin
  \& Romanowsky}{Danieli
  et~al.}{2020}]{DanieliDokkumAbrahamConroyEtAl_2020_TipRedGiant_ApJL}
Danieli S.,  van Dokkum P.,  Abraham R.,  Conroy C.,  Dolphin A.~E.,
  Romanowsky A.~J.,  2020, \mn@doi [ApJL] {10.3847/2041-8213/ab8dc4}, 895, L4

\bibitem[\protect\citeauthoryear{Dehnen \& Aly}{Dehnen \&
  Aly}{2012}]{DehnenAly_2012_ImprovingConvergenceSmoothed_MonNotRAstronSoc}
Dehnen W.,  Aly H.,  2012, \mn@doi [Mon Not R Astron Soc]
  {10.1111/j.1365-2966.2012.21439.x}, 425, 1068

\bibitem[\protect\citeauthoryear{Dekel \& Silk}{Dekel \&
  Silk}{1986}]{DekelSilk_1986_OriginDwarfGalaxies_TheAstrophysicalJournal}
Dekel A.,  Silk J.,  1986, \mn@doi [The Astrophysical Journal]
  {10.1086/164050}, 303, 39

\bibitem[\protect\citeauthoryear{Di~Cintio, Brook, Dutton, Macci{\`o}, Obreja
  \& Dekel}{Di~Cintio
  et~al.}{2017}]{DiCintioBrookDuttonMaccioEtAl_2017_NIHAOXIFormation_MonthlyNoticesoftheRoyalAstronomicalSociety:Letters}
Di~Cintio A.,  Brook C.~B.,  Dutton A.~A.,  Macci{\`o} A.~V.,  Obreja A.,
  Dekel A.,  2017, \mn@doi [Monthly Notices of the Royal Astronomical Society:
  Letters] {10.1093/mnrasl/slw210}, 466, L1

\bibitem[\protect\citeauthoryear{Ferrer \& Hunter}{Ferrer \&
  Hunter}{2013}]{FerrerHunter_2013_ImpactPhasespaceDensity_J.Cosmol.Astropart.Phys.}
Ferrer F.,  Hunter D.~R.,  2013, \mn@doi [J. Cosmol. Astropart. Phys.]
  {10.1088/1475-7516/2013/09/005}, 2013, 005

\bibitem[\protect\citeauthoryear{Gerritsen}{Gerritsen}{1997}]{Gerritsen_1997_StarFormationInterstellar_Ph.D.Thesis}
Gerritsen J. P.~E.,  1997, PhD thesis, \url
  {https://ui.adsabs.harvard.edu/abs/1997PhDT........19G}

\bibitem[\protect\citeauthoryear{Gingold \& Monaghan}{Gingold \&
  Monaghan}{1977}]{GingoldMonaghan_1977_SmoothedParticleHydrodynamics_MonNotRAstronSoc}
Gingold R.~A.,  Monaghan J.~J.,  1977, \mn@doi [Mon Not R Astron Soc]
  {10.1093/mnras/181.3.375}, 181, 375

\bibitem[\protect\citeauthoryear{Guo et~al.,}{Guo
  et~al.}{2020}]{GuoHuZhengLiaoEtAl_2020_FurtherEvidencePopulation_NatAstron}
Guo Q.,  et~al., 2020, \mn@doi [Nat Astron] {10.1038/s41550-019-0930-9}, 4, 246

\bibitem[\protect\citeauthoryear{Hayashi \& Inoue}{Hayashi \&
  Inoue}{2018}]{HayashiInoue_2018_EffectsMassModels_MNRAS}
Hayashi K.,  Inoue S.,  2018, \mn@doi [MNRAS] {10.1093/mnrasl/sly162}, 481, L59

\bibitem[\protect\citeauthoryear{Hunter et~al.,}{Hunter
  et~al.}{2012}]{HunterFicut-VicasAshleyBrinksEtAl_2012_LITTLETHINGS_AJ}
Hunter D.~A.,  et~al., 2012, \mn@doi [AJ] {10.1088/0004-6256/144/5/134}, 144,
  134

\bibitem[\protect\citeauthoryear{Ishiyama \& Ando}{Ishiyama \&
  Ando}{2020}]{IshiyamaAndo_2020_AbundanceStructureSubhaloes_MonthlyNoticesoftheRoyalAstronomicalSociety}
Ishiyama T.,  Ando S.,  2020, \mn@doi [Monthly Notices of the Royal
  Astronomical Society] {10.1093/mnras/staa069}, 492, 3662

\bibitem[\protect\citeauthoryear{Ishiyama et~al.,}{Ishiyama
  et~al.}{2021}]{IshiyamaPradaKlypinSinhaEtAl_2021_UchuuSimulationsData_MonthlyNoticesoftheRoyalAstronomicalSociety}
Ishiyama T.,  et~al., 2021, \mn@doi [Monthly Notices of the Royal Astronomical
  Society] {10.1093/mnras/stab1755}, 506, 4210

\bibitem[\protect\citeauthoryear{Iwasawa, Tanikawa, Hosono, Nitadori, Muranushi
   \& Makino}{Iwasawa
  et~al.}{2016}]{IwasawaTanikawaHosonoNitadoriEtAl_2016_ImplementationPerformanceFDPS_PublAstronSocJpnNihonTenmonGakkai}
Iwasawa M.,  Tanikawa A.,  Hosono N.,  Nitadori K.,  Muranushi T.,   Makino J.,
   2016, \mn@doi [Publ Astron Soc Jpn Nihon Tenmon Gakkai]
  {10.1093/pasj/psw053}, 68

\bibitem[\protect\citeauthoryear{Katz}{Katz}{1992}]{Katz_1992_DissipationalGalaxyFormation_ApJ}
Katz N.,  1992, \mn@doi [ApJ] {10.1086/171366}, 391, 502

\bibitem[\protect\citeauthoryear{Klypin, Kravtsov, Valenzuela  \& Prada}{Klypin
  et~al.}{1999}]{KlypinKravtsovValenzuelaPrada_1999_WhereAreMissing_ApJ}
Klypin A.,  Kravtsov A.~V.,  Valenzuela O.,   Prada F.,  1999, \mn@doi [ApJ]
  {10.1086/307643}, 522, 82

\bibitem[\protect\citeauthoryear{Koda, Yagi, Yamanoi  \& Komiyama}{Koda
  et~al.}{2015}]{KodaYagiYamanoiKomiyama_2015_APPROXIMATHOUSANDULTRADIFFUSE_ApJL}
Koda J.,  Yagi M.,  Yamanoi H.,   Komiyama Y.,  2015, \mn@doi [ApJL]
  {10.1088/2041-8205/807/1/L2}, 807, L2

\bibitem[\protect\citeauthoryear{Lee, Shin  \& Kim}{Lee
  et~al.}{2021}]{LeeShinKim_2021_DarkMatterDeficient_ApJL}
Lee J.,  Shin E.-j.,   Kim J.-h.,  2021, \mn@doi [ApJL]
  {10.3847/2041-8213/ac16e0}, 917, L15

\bibitem[\protect\citeauthoryear{Lucy}{Lucy}{1977}]{Lucy_1977_NumericalApproachTesting_TheAstronomicalJournal}
Lucy L.~B.,  1977, \mn@doi [The Astronomical Journal] {10.1086/112164}, 82,
  1013

\bibitem[\protect\citeauthoryear{Madau, Lupi, Diemand, Burkert  \& Lin}{Madau
  et~al.}{2020}]{MadauLupiDiemandBurkertEtAl_2020_GlobularClusterFormation_ApJ}
Madau P.,  Lupi A.,  Diemand J.,  Burkert A.,   Lin D. N.~C.,  2020, \mn@doi
  [ApJ] {10.3847/1538-4357/ab66c6}, 890, 18

\bibitem[\protect\citeauthoryear{Mancera~Pi{\~n}a et~al.,}{Mancera~Pi{\~n}a
  et~al.}{2019}]{ManceraPinaFraternaliAdamsMarascoEtAl_2019_BaryonicTullyFisher_ApJL}
Mancera~Pi{\~n}a P.~E.,  et~al., 2019, \mn@doi [ApJL]
  {10.3847/2041-8213/ab40c7}, 883, L33

\bibitem[\protect\citeauthoryear{Mancera~Pi{\~n}a et~al.,}{Mancera~Pi{\~n}a
  et~al.}{2020}]{ManceraPinaFraternaliOmanAdamsEtAl_2020_RobustKinematicsGasrich_MNRAS}
Mancera~Pi{\~n}a P.~E.,  et~al., 2020, \mn@doi [MNRAS]
  {10.1093/mnras/staa1256}, 495, 3636

\bibitem[\protect\citeauthoryear{Mancera~Pi{\~n}a, Fraternali, Oosterloo,
  Adams, Oman  \& Leisman}{Mancera~Pi{\~n}a
  et~al.}{2022}]{ManceraPinaFraternaliOosterlooAdamsEtAl_2022_NoNeedDark_MonthlyNoticesoftheRoyalAstronomicalSociety}
Mancera~Pi{\~n}a P.~E.,  Fraternali F.,  Oosterloo T.,  Adams E. A.~K.,  Oman
  K.~A.,   Leisman L.,  2022, \mn@doi [Monthly Notices of the Royal
  Astronomical Society] {10.1093/mnras/stab3491}, 512, 3230

\bibitem[\protect\citeauthoryear{Marinacci et~al.,}{Marinacci
  et~al.}{2018}]{MarinacciVogelsbergerPakmorTorreyEtAl_2018_FirstResultsIllustrisTNG_MonthlyNoticesoftheRoyalAstronomicalSociety}
Marinacci F.,  et~al., 2018, \mn@doi [Monthly Notices of the Royal Astronomical
  Society] {10.1093/mnras/sty2206}, 480, 5113

\bibitem[\protect\citeauthoryear{McConnachie}{McConnachie}{2012}]{McConnachie_2012_OBSERVEDPROPERTIESDWARF_AJ}
McConnachie A.~W.,  2012, \mn@doi [AJ] {10.1088/0004-6256/144/1/4}, 144, 4

\bibitem[\protect\citeauthoryear{Miki \& Umemura}{Miki \&
  Umemura}{2018}]{MikiUmemura_2018_MAGIManycomponentGalaxy_MonNotRAstronSoc}
Miki Y.,  Umemura M.,  2018, \mn@doi [Mon Not R Astron Soc]
  {10.1093/mnras/stx3327}, 475, 2269

\bibitem[\protect\citeauthoryear{Monaghan}{Monaghan}{1997}]{Monaghan_1997_SPHRiemannSolvers_JournalofComputationalPhysics}
Monaghan J.~J.,  1997, \mn@doi [Journal of Computational Physics]
  {10.1006/jcph.1997.5732}, 136, 298

\bibitem[\protect\citeauthoryear{Monelli \& Trujillo}{Monelli \&
  Trujillo}{2019}]{MonelliTrujillo_2019_TRGBDistanceSecond_ApJL}
Monelli M.,  Trujillo I.,  2019, \mn@doi [ApJL] {10.3847/2041-8213/ab2fd2},
  880, L11

\bibitem[\protect\citeauthoryear{Montes, {Infante-Sainz}, {Madrigal-Aguado},
  Rom{\'a}n, Monelli, Borlaff  \& Trujillo}{Montes
  et~al.}{2020}]{MontesInfante-SainzMadrigal-AguadoRomanEtAl_2020_GalaxyMissingDark_ApJ}
Montes M.,  {Infante-Sainz} R.,  {Madrigal-Aguado} A.,  Rom{\'a}n J.,  Monelli
  M.,  Borlaff A.~S.,   Trujillo I.,  2020, \mn@doi [ApJ]
  {10.3847/1538-4357/abc340}, 904, 114

\bibitem[\protect\citeauthoryear{Montes, Trujillo, {Infante-Sainz}, Monelli  \&
  Borlaff}{Montes
  et~al.}{2021}]{MontesTrujilloInfante-SainzMonelliEtAl_2021_DiskNoSignatures_ApJ}
Montes M.,  Trujillo I.,  {Infante-Sainz} R.,  Monelli M.,   Borlaff A.~S.,
  2021, \mn@doi [ApJ] {10.3847/1538-4357/ac0d55}, 919, 56

\bibitem[\protect\citeauthoryear{Moore, Ghigna, Governato, Lake, Quinn, Stadel
  \& Tozzi}{Moore
  et~al.}{1999}]{MooreGhignaGovernatoLakeEtAl_1999_DarkMatterSubstructure_ApJ}
Moore B.,  Ghigna S.,  Governato F.,  Lake G.,  Quinn T.,  Stadel J.,   Tozzi
  P.,  1999, \mn@doi [ApJ] {10.1086/312287}, 524, L19

\bibitem[\protect\citeauthoryear{Mori, Yoshii, Tsujimoto  \& Nomoto}{Mori
  et~al.}{1997}]{MoriYoshiiTsujimotoNomoto_1997_EvolutionDwarfGalaxies_ApJ}
Mori M.,  Yoshii Y.,  Tsujimoto T.,   Nomoto K.,  1997, \mn@doi [ApJ]
  {10.1086/310547}, 478, L21

\bibitem[\protect\citeauthoryear{Mori, Yoshii  \& Nomoto}{Mori
  et~al.}{1999}]{MoriYoshiiNomoto_1999_DissipativeProcessMechanism_ApJ}
Mori M.,  Yoshii Y.,   Nomoto K.,  1999, \mn@doi [ApJ] {10.1086/306724}, 511,
  585

\bibitem[\protect\citeauthoryear{M{\"u}ller et~al.,}{M{\"u}ller
  et~al.}{2019}]{MullerRichRomanYildizEtAl_2019_TidalTaleDetection_A&A}
M{\"u}ller O.,  et~al., 2019, \mn@doi [A\&A] {10.1051/0004-6361/201935463},
  624, L6

\bibitem[\protect\citeauthoryear{Naiman et~al.,}{Naiman
  et~al.}{2018}]{NaimanPillepichSpringelRamirez-RuizEtAl_2018_FirstResultsIllustrisTNG_MonthlyNoticesoftheRoyalAstronomicalSociety}
Naiman J.~P.,  et~al., 2018, \mn@doi [Monthly Notices of the Royal Astronomical
  Society] {10.1093/mnras/sty618}, 477, 1206

\bibitem[\protect\citeauthoryear{Namekata et~al.,}{Namekata
  et~al.}{2018}]{NamekataIwasawaNitadoriTanikawaEtAl_2018_FortranInterfaceLayer_PublicationsoftheAstronomicalSocietyofJapan}
Namekata D.,  et~al., 2018, \mn@doi [Publications of the Astronomical Society
  of Japan] {10.1093/pasj/psy062}, 70

\bibitem[\protect\citeauthoryear{Navarro, Frenk  \& White}{Navarro
  et~al.}{1996}]{NavarroFrenkWhite_1996_StructureColdDark_Astrophys.J.}
Navarro J.~F.,  Frenk C.~S.,   White S. D.~M.,  1996, \mn@doi [The
  Astrophysical Journal] {10.1086/177173}, 462, 563

\bibitem[\protect\citeauthoryear{Navarro, Frenk  \& White}{Navarro
  et~al.}{1997}]{NavarroFrenkWhite_1997_UniversalDensityProfile_ApJa}
Navarro J.~F.,  Frenk C.~S.,   White S. D.~M.,  1997, \mn@doi [ApJ]
  {10.1086/304888}, 490, 493

\bibitem[\protect\citeauthoryear{Nelson et~al.,}{Nelson
  et~al.}{2018}]{NelsonPillepichSpringelWeinbergerEtAl_2018_FirstResultsIllustrisTNG_MonthlyNoticesoftheRoyalAstronomicalSociety}
Nelson D.,  et~al., 2018, \mn@doi [Monthly Notices of the Royal Astronomical
  Society] {10.1093/mnras/stx3040}, 475, 624

\bibitem[\protect\citeauthoryear{Nelson et~al.,}{Nelson
  et~al.}{2019}]{NelsonPillepichSpringelPakmorEtAl_2019_FirstResultsTNG50_MonthlyNoticesoftheRoyalAstronomicalSociety}
Nelson D.,  et~al., 2019, \mn@doi [Monthly Notices of the Royal Astronomical
  Society] {10.1093/mnras/stz2306}, 490, 3234

\bibitem[\protect\citeauthoryear{Nusser}{Nusser}{2020}]{Nusser_2020_ScenarioUltradiffuseSatellite_ApJ}
Nusser A.,  2020, \mn@doi [ApJ] {10.3847/1538-4357/ab792c}, 893, 66

\bibitem[\protect\citeauthoryear{Ogiya}{Ogiya}{2018}]{Ogiya_2018_TidalStrippingPossible_MNRAS}
Ogiya G.,  2018, \mn@doi [MNRAS] {10.1093/mnrasl/sly138}, 480, L106

\bibitem[\protect\citeauthoryear{Oh et~al.,}{Oh
  et~al.}{2015}]{OhHunterBrinksElmegreenEtAl_2015_HIGHRESOLUTIONMASSMODELS_AJ}
Oh S.-H.,  et~al., 2015, \mn@doi [AJ] {10.1088/0004-6256/149/6/180}, 149, 180

\bibitem[\protect\citeauthoryear{Oku, Tomida, Nagamine, Shimizu  \& Cen}{Oku
  et~al.}{2022}]{OkuTomidaNagamineShimizuEtAl_2022_OsakaFeedbackModel_ApJS}
Oku Y.,  Tomida K.,  Nagamine K.,  Shimizu I.,   Cen R.,  2022, \mn@doi [ApJS]
  {10.3847/1538-4365/ac77ff}, 262, 9

\bibitem[\protect\citeauthoryear{Otaki \& Mori}{Otaki \&
  Mori}{2022a}]{OtakiMori_2022_StudyGalaxyCollisions_Comput.Sci.ItsAppl.-ICCSA2022Workshop}
Otaki K.,  Mori M.,  2022a, in Gervasi O.,  Murgante B.,  Misra S.,  Rocha A.
  M. A.~C.,   Garau C.,  eds, Computational {{Science}} and {{Its
  Applications}} \textendash{} {{ICCSA}} 2022 {{Workshops}}. Lecture {{Notes}}
  in {{Computer Science}}.
{Springer International Publishing}, {Cham}, pp 373--387,
  \mn@doi{10.1007/978-3-031-10562-3\_27}

\bibitem[\protect\citeauthoryear{Otaki \& Mori}{Otaki \&
  Mori}{2022b}]{OtakiMori_2022_FormationDarkmatterdeficientGalaxies_J.Phys.:Conf.Ser.}
Otaki K.,  Mori M.,  2022b, \mn@doi [J. Phys.: Conf. Ser.]
  {10.1088/1742-6596/2207/1/012049}, 2207, 012049

\bibitem[\protect\citeauthoryear{Otaki \& Mori}{Otaki \&
  Mori}{2023}]{OtakiMori_2023_CollisioninducedFormationDarkmatterdeficient_IAUSymp.}
Otaki K.,  Mori M.,  2023, \mn@doi [IAU Symposium] {10.1017/S1743921322004562},
  373, 147

\bibitem[\protect\citeauthoryear{Paudel, Smith, Yoon, {Calder{\'o}n-Castillo}
  \& Duc}{Paudel
  et~al.}{2018}]{PaudelSmithYoonCalderon-CastilloEtAl_2018_CatalogMergingDwarf_ApJS}
Paudel S.,  Smith R.,  Yoon S.~J.,  {Calder{\'o}n-Castillo} P.,   Duc P.-A.,
  2018, \mn@doi [ApJS] {10.3847/1538-4365/aad555}, 237, 36

\bibitem[\protect\citeauthoryear{Pillepich et~al.,}{Pillepich
  et~al.}{2018}]{PillepichNelsonHernquistSpringelEtAl_2018_FirstResultsIllustrisTNG_MonNotRAstronSoc}
Pillepich A.,  et~al., 2018, \mn@doi [Mon Not R Astron Soc]
  {10.1093/mnras/stx3112}, 475, 648

\bibitem[\protect\citeauthoryear{Pillepich et~al.,}{Pillepich
  et~al.}{2019}]{PillepichNelsonSpringelPakmorEtAl_2019_FirstResultsTNG50_MonthlyNoticesoftheRoyalAstronomicalSociety}
Pillepich A.,  et~al., 2019, \mn@doi [Monthly Notices of the Royal Astronomical
  Society] {10.1093/mnras/stz2338}, 490, 3196

\bibitem[\protect\citeauthoryear{Poulain et~al.,}{Poulain
  et~al.}{2022}]{PoulainMarleauHabasDucEtAl_2022_HIObservationsMATLAS_A&A}
Poulain M.,  et~al., 2022, \mn@doi [A\&A] {10.1051/0004-6361/202142012}, 659,
  A14

\bibitem[\protect\citeauthoryear{Prada, Klypin, Cuesta, {Betancort-Rijo}  \&
  Primack}{Prada
  et~al.}{2012}]{PradaKlypinCuestaBetancort-RijoEtAl_2012_HaloConcentrationsStandard_MonNotRAstronSoc}
Prada F.,  Klypin A.~A.,  Cuesta A.~J.,  {Betancort-Rijo} J.~E.,   Primack J.,
  2012, \mn@doi [Mon Not R Astron Soc] {10.1111/j.1365-2966.2012.21007.x}, 423,
  3018

\bibitem[\protect\citeauthoryear{Salpeter}{Salpeter}{1955}]{Salpeter_1955_LuminosityFunctionStellar_TheAstrophysicalJournal}
Salpeter E.~E.,  1955, \mn@doi [The Astrophysical Journal] {10.1086/145971},
  121, 161

\bibitem[\protect\citeauthoryear{Shen et~al.,}{Shen
  et~al.}{2021}]{ShenDanieliDokkumAbrahamEtAl_2021_TipRedGiant_ApJL}
Shen Z.,  et~al., 2021, \mn@doi [ApJL] {10.3847/2041-8213/ac0335}, 914, L12

\bibitem[\protect\citeauthoryear{Shimizu, Todoroki, Yajima  \&
  Nagamine}{Shimizu
  et~al.}{2019}]{ShimizuTodorokiYajimaNagamine_2019_OsakaFeedbackModel_MonNotRAstronSoc}
Shimizu I.,  Todoroki K.,  Yajima H.,   Nagamine K.,  2019, \mn@doi [Mon Not R
  Astron Soc] {10.1093/mnras/stz098}, 484, 2632

\bibitem[\protect\citeauthoryear{Shin, Jung, Kwon, Kim, Lee, Jo  \& Oh}{Shin
  et~al.}{2020}]{ShinJungKwonKimEtAl_2020_DarkMatterDeficient_ApJ}
Shin E.-j.,  Jung M.,  Kwon G.,  Kim J.-h.,  Lee J.,  Jo Y.,   Oh B.~K.,  2020,
  \mn@doi [ApJ] {10.3847/1538-4357/aba434}, 899, 25

\bibitem[\protect\citeauthoryear{Silk}{Silk}{2019}]{Silk_2019_UltradiffuseGalaxiesDark_MNRAS}
Silk J.,  2019, \mn@doi [MNRAS] {10.1093/mnrasl/slz090}, 488, L24

\bibitem[\protect\citeauthoryear{Springel \& Hernquist}{Springel \&
  Hernquist}{2002}]{SpringelHernquist_2002_CosmologicalSmoothedParticle_MonthlyNoticesoftheRoyalAstronomicalSociety}
Springel V.,  Hernquist L.,  2002, \mn@doi [Monthly Notices of the Royal
  Astronomical Society] {10.1046/j.1365-8711.2002.05445.x}, 333, 649

\bibitem[\protect\citeauthoryear{Springel et~al.,}{Springel
  et~al.}{2018}]{SpringelPakmorPillepichWeinbergerEtAl_2018_FirstResultsIllustrisTNG_MonthlyNoticesoftheRoyalAstronomicalSociety}
Springel V.,  et~al., 2018, \mn@doi [Monthly Notices of the Royal Astronomical
  Society] {10.1093/mnras/stx3304}, 475, 676

\bibitem[\protect\citeauthoryear{Stierwalt, Besla, Patton, Johnson,
  Kallivayalil, Putman, Privon  \& Ross}{Stierwalt
  et~al.}{2015}]{StierwaltBeslaPattonJohnsonEtAl_2015_TiNyTITANSROLE_ApJ}
Stierwalt S.,  Besla G.,  Patton D.,  Johnson K.,  Kallivayalil N.,  Putman M.,
   Privon G.,   Ross G.,  2015, \mn@doi [ApJ] {10.1088/0004-637X/805/1/2}, 805,
  2

\bibitem[\protect\citeauthoryear{Sutherland \& Dopita}{Sutherland \&
  Dopita}{2017}]{SutherlandDopita_2017_EffectsPreionizationRadiative_ApJS}
Sutherland R.~S.,  Dopita M.~A.,  2017, \mn@doi [ApJS]
  {10.3847/1538-4365/aa6541}, 229, 34

\bibitem[\protect\citeauthoryear{Sutherland, Dopita, Binette  \&
  Groves}{Sutherland
  et~al.}{2018}]{SutherlandDopitaBinetteGroves_2018_MAPPINGSAstrophysicalPlasma_AstrophysicsSourceCodeLibrary}
Sutherland R.,  Dopita M.,  Binette L.,   Groves B.,  2018, Astrophysics Source
  Code Library, p. ascl:1807.005

\bibitem[\protect\citeauthoryear{Thacker \& Couchman}{Thacker \&
  Couchman}{2000}]{ThackerCouchman_2000_ImplementingFeedbackSimulations_Astrophys.J.}
Thacker R.~J.,  Couchman H. M.~P.,  2000, \mn@doi [The Astrophysical Journal]
  {10.1086/317828}, 545, 728

\bibitem[\protect\citeauthoryear{Townsend}{Townsend}{2009}]{Townsend_2009_EXACTINTEGRATIONSCHEME_ApJS}
Townsend R. H.~D.,  2009, \mn@doi [ApJS] {10.1088/0067-0049/181/2/391}, 181,
  391

\bibitem[\protect\citeauthoryear{Trujillo et~al.,}{Trujillo
  et~al.}{2019}]{TrujilloBeasleyBorlaffCarrascoEtAl_2019_Distance13Mpc_MNRAS}
Trujillo I.,  et~al., 2019, \mn@doi [MNRAS] {10.1093/mnras/stz771}, 486, 1192

\bibitem[\protect\citeauthoryear{Wasserman, Romanowsky, Brodie, van Dokkum,
  Conroy, Abraham, Cohen  \& Danieli}{Wasserman
  et~al.}{2018}]{WassermanRomanowskyBrodieDokkumEtAl_2018_DeficitDarkMatter_ApJLa}
Wasserman A.,  Romanowsky A.~J.,  Brodie J.,  van Dokkum P.,  Conroy C.,
  Abraham R.,  Cohen Y.,   Danieli S.,  2018, \mn@doi [ApJL]
  {10.3847/2041-8213/aad779}, 863, L15

\bibitem[\protect\citeauthoryear{Wendland}{Wendland}{1995}]{Wendland_1995_PiecewisePolynomialPositive_AdvComputMath}
Wendland H.,  1995, \mn@doi [Adv Comput Math] {10.1007/BF02123482}, 4, 389

\bibitem[\protect\citeauthoryear{Yang, Yu  \& An}{Yang
  et~al.}{2020}]{YangYuAn_2020_SelfInteractingDarkMatter_Phys.Rev.Lett.b}
Yang D.,  Yu H.-B.,   An H.,  2020, \mn@doi [Phys. Rev. Lett.]
  {10.1103/PhysRevLett.125.111105}, 125, 111105

\bibitem[\protect\citeauthoryear{Zhu, Hernquist  \& Li}{Zhu
  et~al.}{2015}]{ZhuHernquistLi_2015_NUMERICALCONVERGENCESMOOTHED_ApJ}
Zhu Q.,  Hernquist L.,   Li Y.,  2015, \mn@doi [ApJ]
  {10.1088/0004-637X/800/1/6}, 800, 6

\bibitem[\protect\citeauthoryear{van Dokkum, Abraham, Merritt, Zhang, Geha  \&
  Conroy}{van Dokkum
  et~al.}{2015}]{DokkumAbrahamMerrittZhangEtAl_2015_FORTYSEVENMILKYWAYSIZED_ApJL}
van Dokkum P.~G.,  Abraham R.,  Merritt A.,  Zhang J.,  Geha M.,   Conroy C.,
  2015, \mn@doi [ApJL] {10.1088/2041-8205/798/2/L45}, 798, L45

\bibitem[\protect\citeauthoryear{{van Dokkum} et~al.,}{{van Dokkum}
  et~al.}{2018}]{vanDokkumDanieliCohenMerrittEtAl_2018_GalaxyLackingDark_Nature}
{van Dokkum} P.,  et~al., 2018, \mn@doi [Nature] {10.1038/nature25767}, 555,
  629

\bibitem[\protect\citeauthoryear{van Dokkum, Danieli, Abraham, Conroy  \&
  Romanowsky}{van Dokkum
  et~al.}{2019}]{DokkumDanieliAbrahamConroyEtAl_2019_SecondGalaxyMissing_ApJL}
van Dokkum P.,  Danieli S.,  Abraham R.,  Conroy C.,   Romanowsky A.~J.,  2019,
  \mn@doi [ApJL] {10.3847/2041-8213/ab0d92}, 874, L5

\bibitem[\protect\citeauthoryear{{van Dokkum} et~al.,}{{van Dokkum}
  et~al.}{2022a}]{vanDokkumShenKeimTrujillo-GomezEtAl_2022_TrailDarkmatterfreeGalaxies_Nature}
{van Dokkum} P.,  et~al., 2022a, \mn@doi [Nature] {10.1038/s41586-022-04665-6},
  605, 435

\bibitem[\protect\citeauthoryear{{van Dokkum} et~al.,}{{van Dokkum}
  et~al.}{2022b}]{vanDokkumShenRomanowskyAbrahamEtAl_2022_MonochromaticGlobularClusters_Astrophys.J.}
{van Dokkum} P.,  et~al., 2022b, \mn@doi [The Astrophysical Journal]
  {10.3847/2041-8213/ac94d6}, 940, L9

\makeatother
\end{thebibliography}







\bsp	
\label{lastpage}
\end{document}